\documentclass[aps,prb,twocolumn,floatfix,superscriptaddress,10pt]{revtex4-2}
\usepackage[utf8]{inputenc}
\usepackage[T1]{fontenc}
\usepackage{lmodern}
\usepackage[pdftex,colorlinks=true,bookmarksnumbered,bookmarksopen=true,bookmarksopenlevel=-1]{hyperref}
\usepackage{amsmath}
\usepackage{amssymb}
\usepackage{graphicx}
\usepackage{bm}
\usepackage{color}
\usepackage{hyperref}
\usepackage{upgreek}
\usepackage{scalerel}
\usepackage{pgffor}
\usepackage{pdfpages}
\usepackage{float}
\usepackage{physics} 
\usepackage{lscape}   
\usepackage{comment}  
\makeatletter
\AtBeginDocument{\let\LS@rot\@undefined}
\makeatother

\usepackage{silence}
\makeindex

\newcommand{\Eexp}[1]{ \ensuremath{ \textrm{e}^{ #1} } }
\newcommand{\BoldVec}[1]{ \ensuremath{ {\boldsymbol{#1}} } }

\begin{document}

\title{%
	Efficient numerical method for evaluating normal and anomalous time-domain equilibrium Green's functions in inhomogeneous systems 
}
\author{Tomas L\"{o}thman}
	\affiliation{Department of Physics and Astronomy, Uppsala University, Box 516, S-751 20 Uppsala, Sweden}
\author{Christopher Triola} 
	\affiliation{Los Alamos National Laboratory, Los Alamos, New Mexico 87544, USA}
\author{Jorge Cayao} 
	\email[Corresponding author: ]{jorge.cayao@physics.uu.se}
	\affiliation{Department of Physics and Astronomy, Uppsala University, Box 516, S-751 20 Uppsala, Sweden}
\author{Annica M. Black-Schaffer}
	\affiliation{Department of Physics and Astronomy, Uppsala University, Box 516, S-751 20 Uppsala, Sweden}

\date{\today}
\begin{abstract}
	In this work we develop EPOCH: Equilibrium Propagator by Orthogonal polynomial CHain, a computationally  efficient method to calculate the time-dependent equilibrium Green's functions, including the anomalous Green's functions of superconductors, to capture the time-evolution in large inhomogeneous systems.	The EPOCH method generalizes  the Chebyshev wave-packet propagation method from quantum chemistry and efficiently incorporates the Fermi-Dirac statistics that is needed for equilibrium quantum condensed matter systems.	The computational cost of EPOCH scales only linearly in the system degrees of freedom, generating an extremely efficient algorithm also for very large systems. 
	We demonstrate the power of the EPOCH method by calculating the time-evolution of  an excitation near a superconductor-normal metal interface in two and three dimensions, capturing transmission as well as normal and Andreev reflections.
	\end{abstract}\maketitle

\section{Introduction}
\label{sec:intro}
	The study of time-evolution in quantum mechanical systems provides fundamental insights into the system's underlying structure, allowing for the development of applications harnessing dynamical quantum effects, with prominent examples in quantum information processing \cite{Baeuerle2018, Bertoni2000, Kloeffel2013, Brif2010}. Notably, even quantum systems without any external time-dependent drive usually present a highly complex dynamical behavior, exhibited in its time-dependent correlation functions without the need to go out of equilibrium. The dynamical behaviors range from the transmission and reflectance across interfaces to more general quantum transport properties, including also quantum electronics with coherent single electron excitations in solids \cite{Feve1169, PhysRevLett.111.216807,Dasenbrook_2016, Baeuerle2018}. In superconductors, another remarkable time-dependent phenomenon occurs when electrons pair at unequal times, enabling odd-frequency pairing, present in for example superconductor-ferromagnetic and multiband systems \cite{PhysRevLett.86.4096,RevModPhys.77.1321,PhysRevB.88.104514,Nagaosa12,RevModPhys.91.045005,cayao2019odd,triola2020role}. Moreover, through the fluctuation-dissipation theorem, equilibrium correlations even predict the response to time-dependent external perturbations \cite{Zwanzig1965,Kubo1966,mahan2000}. 			
	
	At the same time, many systems with highly non-trivial quantum dynamics also lack translation invariance, such as those dominated by the presence of junctions, interfaces, edges, or disorder. Therefore, these systems also see a dramatic growth in the number of degrees of freedom that must be treated simultaneously. This fact makes the use of analytical methods, as well as brute force eigenvalue diagonalization, completely unrealistic, even for systems with a noninteracting quasiparticle description and in equilibrium. Faced by this very challenge, we seek a method that efficiently and accurately compute time-domain properties with a low computational cost even as the number of degrees of freedom proliferate.
	
	To extract interesting dynamical properties, the correct objects are usually the time-domain Green's functions, encoding the physical observables of any system. Fortunately, for any generic non-interacting system, all higher order Green's functions reduce to the computation of the single particle Green's functions by virtue of Wick's theorem \cite{mahan2000}. The Green's functions are also by themselves inherently interesting objects, representing the direct physical probability amplitude of finding a particle at a point  in space-time $(x_2,t_2)$, after an earlier insertion at $(x_1,t_1)$. In superconductors, where particle number is not conserved, the inserted particle may later be found as a hole, and the amplitude for this conversion process is properly handled by the anomalous Green's functions. Here, as an example, the unequal time anomalous Green's functions may display exotic odd-frequency dynamical pairing \cite{RevModPhys.91.045005,cayao2019odd}, and are therefore also of large interest.
	
	In terms of evaluating any physical observables in the time-domain, we first note that, for pure quantum states, the time-domain naturally lends itself to a step-by-step propagation in time without the need of any costly diagonalization. Simply because the Hamiltonian $H$ is precisely the time-evolution generator in any quantum system. Therefore, the time-dependence of a pure quantum state, i.e.~single wave-packet propagation, is directly achievable with only matrix-vector multiplication; for instance, by Taylor expanding the unitary time evolution operator $U(t) = e^{- i H t}$, by naive finite differences of the Schr\"{o}dinger differential equation, or by somewhat more involved algorithms \cite{Iitaka1994}.
	
	The single wave-packet propagation methods are vastly improved, both in terms of accuracy and efficiency (exponential over power law convergence), by the Chebyshev method of time-evolution \cite{Tal-Ezer1984}, a method that is already firmly established in quantum chemistry and molecular dynamics \cite{Kosloff1988,Leforestier1991,Kosloff1994,Garraway1995}. The basis of the Chebyshev time-evolution method is the expansion of the evolution operator in terms of Bessel functions $J_{n}(t)$ and the Chebyshev polynomials $T_n(x)$ as: 
	$U(t) = \sum_{n=0}^{\infty} (2 - \delta_{n 0}) (-i)^{n} J_{n}( \tilde{t} ) T_{n}( \tilde{H} )$
	\footnote{		
		The expansion follows from the well-known Jacobi-Anger expansion \unexpanded{$e^{i z \cos \theta} = \sum_{n=0}^{\infty} i^{n} (2 - \delta_{n 0}) J_{n}(z) \cos (n \theta)$} and the definition of the  Chebyshev polynomials  $T_n(x) = \cos ( n \arccos (x ) )$
	}, where $\tilde{H}=H/\left\| H \right\|$ and $\tilde{t}=\left\| H \right\| t$ to fit the standard domain of $T_n(x)$. The resulting series has the advantageous property that it converges rapidly, because for $\tilde{t} \lesssim n$ the Bessel $J_{n}(\tilde{t})$ is exponentially suppressed. Therefore, before a given finite time $\tilde{t}$, it is only the first $n$ terms that contribute significantly. A second underpinning of the method's efficiency is that at each expansion order, the next order is obtained via a recursive relationship connecting the $T_{n}(\tilde{H})$ polynomials of neighboring degree. This limits the computational cost to just one additional (sparse) multiplication by the Hamiltonian $\tilde{H}$ per expansion order. As a result, the Chebyshev wave-packet propagation method achieves both accuracy and stability with a low computational cost.

	In a broader context, the efficiency of the Chebyshev wave-packet propagation method relates directly to the advantageous properties of the underlying Chebyshev polynomials $T_n(x)$, which, like all orthogonal polynomials, form a complete basis set. It is therefore not surprising that orthogonal polynomials have not just been found to be useful for wave-function propagation, but also have a long history as a versatile technique in approximation theory \cite{Cheney2000} and also applied to electronic structure methods, where they have developed into a prominent linear-scaling method known as the kernel polynomial method (KPM) \cite{Goedecker1999, Weise2006}. This method has notably been used to compute spectral functions and expectation values in very large systems \cite{Silver1996, PhysRevB.53.12733} and underpin several new software packages \cite{Fan2018,Bjrnson2019,Joao2020,Moldovan2020}. Despite this, it was only recently that KPM was extended to also tackle superconducting systems \cite{PhysRevLett.105.167006, Nagai2012}, but still not in the time-domain, which is necessary when studying, for instance, time evolution.
	 
While the Chebyshev wave-packet propagation method has already been highly successful to evaluate observables in the time-domain, a fundamental limitation for condensed matter system is that it is only applicable to pure quantum states, i.e.~single electrons, and not to many electron states, i.e.~ mixed states. Thus it cannot evaluate, for instance, thermodynamic expectation values, which  are central concepts in condensed matter. Therefore, in order to achieve the same efficiency and accuracy as the Chebyshev wave-packet propagation method also in condensed matter systems, we need to incorporate Fermi-Dirac statistics alongside the time-evolution.
	
	In this work, we develop the Equilibrium Propagator by Orthogonal polynomial CHain, or EPOCH, method, to efficiently calculate the time-domain equilibrium Green's functions for both normal and anomalous correlations in generic quantum condensed matter systems. Generalizing the Chebyshev wave-packet propagation method from quantum chemistry, the EPOCH method efficiently incorporates the Fermi-Dirac statistics that is needed for the equilibrium physics of quantum condensed matter systems.
	
	To derive the EPOCH method, we first strategically depart from the Chebyshev method and instead use the Legendre polynomials, since they in contrast allow us to derive analytical expressions for key quantities in the time-domain. This changes our reference starting point, from the expansion of the evolution operator $U(t)$ in terms of the Chebyshev polynomials $T_n(x)$ to an analogous expansion in terms of the Legendre polynomials $P_{n}(x)$ (see \cite{Kosloff1994}, Eq.~(7.12))
		\begin{equation}				
			\label{Eq:LegendreEvolution}
			U(t) = \sum_{n=0}^{\infty}(-i)^{n} (2 n + 1) j_{n}(\tilde{t}) P_{n}(\tilde{H})\,,
		\end{equation}
	where $j_{n}(\tilde{t})$ are the spherical Bessel functions 
		\footnote{%
			This expansion follows from the well-known plane wave expansion \unexpanded{$
			e^{i \mathbf{k} \cdot \mathbf{r}}
			=
			\sum_{n=0}^{\infty} i^{n} (2 n + 1) j_{n}(k r) P_{n}( \hat{\mathbf{k}} \cdot \hat{\mathbf{r}} )
			$}
		.}.
	Next, in order to move beyond the wave-packet propagation and towards a Green's function formalism, we in Section \ref{sec:BdG} first review the standard Bogoliubov-de Gennes (BdG) formalism  for a generic time-independent Hamiltonian $H $ with an emphasis on the time-domain \cite{mahan2000,zagoskin}, since this formalism is applicable to systems both with and without superconductivity. Thereafter, we show how, at any inverse temperature $\beta=1/T$, the thermal equilibrium Green's functions (the lesser $\mathcal{G}^{<} (t)$, greater $\mathcal{G}^{>} (t)$, and anomalous Green's functions $\mathcal{F} (t)$) are all found as the matrix elements of a quantity which we refer to as the Equilibrium Propagator (EP) given by $L_{\beta}(H,t) = \Eexp{ - i H t } F_\beta \left( H \right)$, with $F_\beta \left( H \right)$ being the Fermi-Dirac function. The key component of the EPOCH method then boils down to a general series expansion of the EP, which we derive in Section \ref{sec:OPandEPP}. When specifically applied to the Legendre polynomials $P_n (x)$ in Section \ref{sec:Legendre}, we arrive at,
	\begin{equation}		
		\begin{split}
				\label{Eq:EPPEXP}
				L_{\beta}(H,t) 
				& = 
				\frac{1}{i}
				\begin{pmatrix}
					\mathcal{G}^{<} (t) 			& \mathcal{F} (t) \\
					\mathcal{F}^{\dagger} (t)	& [\mathcal{G}^{>} (t)]^{*}
				\end{pmatrix}
				\\
				& = 
				\frac{1}{2} 
				\sum_{n=0}^{\infty}
				(2 n + 1) 
				(-i)^{n} ( j_{n} (\tilde{t})  + i f^{n}_{\tilde{\beta}} ( \tilde{t} ) ) P_n (\tilde{H})	\,,
			\end{split}
	\end{equation}
	where $\tilde{H}=H/\left\| H \right\|$ and $\tilde{t}=\left\| H \right\| t$ to fit the standard domain of $P_n(x)$. This expansion can be thought of as a generalization to many-body fermionic systems of the Chebyshev wave-packet propagation method already widespread in quantum chemistry \cite{Tal-Ezer1984, Kosloff1988,Leforestier1991,Kosloff1994,Garraway1995} or, likewise, as the series expansion for the time-evolution operator in Eq.~\eqref{Eq:LegendreEvolution}. Within this generalization, the coefficients   $l^{n}_{\tilde{\beta}} (\tilde{t}) = (-i)^{n} ( j_{n} (\tilde{t})  + i f^{n}_{\tilde{\beta}} \left( \tilde{t} \right) )$ receive an additional second part $f^{n}_{\tilde{\beta}} \left( \tilde{t} \right)$ not present in the time-evolution operator in Eq.~\eqref{Eq:LegendreEvolution}, which only has the coefficients $j_{n} (\tilde{t})$. The additional term $f^{n}_{\tilde{\beta}} \left( \tilde{t} \right)$ is a projective mode transient and contains all the temperature dependence and therefore fully encapsulates the Fermi-Dirac function $F_\beta \left( H \right)$. Thus, with just the crucial change of including the projective mode transient, the full quantum statistics is taken into account. The projective mode transients $f^{n}_{\tilde{\beta}} \left( \tilde{t} \right)$ are, as shown in Section~\ref{sec:Legendre}, the solutions to a closed recurrence relationship with an inhomogeneous source term, and we also show how to calculate them with a numerically stable and efficient method.
	
The computational cost in the EPOCH method is kept extremely low as the method scales linearly with the degrees of freedom (system size) and with the longest evolved time, i.e.~running the time evolution for twice as long will only take twice as long to compute, all else being equal. This is because each expansion order of Eq.~\eqref{Eq:EPPEXP} requires only one additional multiplication by $H$ (on very general grounds is a sparse matrix), which follows from the  three-term recursion relationship of the Legendre polynomials. Another strength of the EPOCH method is that any evolved time (or arbitrary set of times) at all temperatures is directly and easily computable.  Moreover, the formalism offers a detailed account of any error introduced by the inevitably finite truncation of the EP series expansion. In Section \ref{sec:Truncation} we establish the accuracy of the expansion, and also show how Gibbs phenomenon, a possible error source due to the sharp Fermi surface in systems without an energy gap, is effectively avoided by going to small but finite temperatures. For easy use and summary, we end the formal development of the EPOCH method in Section \ref{sec:algorithm} by providing a step-by-step outline of the numerical implementation together with the error bounds.
	
	As a demonstration  of the capabilities of the EPOCH method, we first use it in Section \ref{sec:SN} to calculate the particle propagation both across and around a junction between a superconductor and a normal metal in both two and  three dimensions. Despite the very large system sizes involved, especially for the three-dimensional problem, we capture all the transmission and reflection amplitudes, including the Andreev processes.  While these examples are simple, they nonetheless clearly illustrate the capabilities of the EPOCH method, and how it enables efficient computation of the complex quantum dynamics even on a standard laptop.  As a second demonstration, we show in Section \ref{sec:LinearResponse} how the dynamical correlations between observables and the general time-dependent linear response to external probes within the Kubo formalism can be computed directly within the EPOCH method. In this application, a clear advantage of EPOCH is that it automatically gives all time and temperature dependence, without the need to recompute the quantum propagation. EPOCH therefore enables calculation of the response to pulses  probes directly in the time-domain, predicting observables and material properties measured by e.g.~scattering, polarizability, and transport \cite{giuliani2005quantum}.
	
\section{Time-dependent Bogoliubov-de Gennes formalism} \label{sec:BdG}
	To provide a simple starting point and ensure that our approach is applicable to condensed matter systems with or without superconductivity, we review the standard BdG formalism. We consider a fully general quadratic time-independent Hamiltonian $H$ and separate it into two parts $\hat{H} = \hat{H}_{0} + \hat{\Delta}$ where the first part $\hat{H}_{0}$ conserves particle number, thus capturing the normal part, while the second part $\hat{\Delta}$ contains all terms that break the gauge invariance, notably the terms of a superconducting condensate, if present. Such a Hamiltonian models any device, with both a spatially varying normal state or a superconducting order in any part of the system.
	
	Adopting the block Nambu-spinor 
	$ 
		X = 
		\begin{pmatrix}
			c & 	c^\dagger \\
		\end{pmatrix}^T 
	$ of dimension $2N$ for the $N$ degrees of freedom in the system (lattice sites, spin, and orbital degrees of freedom), the Hamiltonian takes, up to a constant shift, the BdG bilinear form  \cite{mahan2000,zagoskin}
	\begin{equation}
		\label{Eq:BdGForm}
		\hat{H}
		=
		X^\dagger H X
		=
		\begin{pmatrix}
			c^\dagger			& 	c			 		\\
		\end{pmatrix}
		\begin{pmatrix}
			H_{0} 					& 	\Delta 		\\
			\Delta^\dagger 	& 	-H^T_{0} 	\\
		\end{pmatrix}
		\begin{pmatrix}
			c 					\\
			c^\dagger 	\\
		\end{pmatrix}\,.
	\end{equation}
	Using the matrix block structure, the eigenvectors of the BdG Hamiltonian can be written as, 
	$
		H 
		\begin{pmatrix}
			 u & v^*		 		\\
		\end{pmatrix}^T  
		= 
		E 
		\begin{pmatrix}
			 u & v^*		 		\\
		\end{pmatrix}^T 
	$, with amplitudes $u$ and $v$.

	The particle-hole symmetry inherent to the BdG Hamiltonian implies that each eigenvector
	also has a symmetry companion state of the opposite energy, 
	$
		H 
		\begin{pmatrix}
			 v & u^*		 		\\
		\end{pmatrix}^T  
		= 
		-E 
		\begin{pmatrix}
			v & u^*		 		\\
		\end{pmatrix}^T
		.
	$	
	Thus, the unitary transformation, $U^\dagger U = I $, with the block structure 
	\[
		U
		=
		\begin{pmatrix}
			u			&	v 		\\
			v^* 	&	u^* 	
		\end{pmatrix},
	\]
	diagonalizes the Hamiltonian:
	$
		\hat{H}
		= 
		\left( X^\dagger U \right) 
		\left( U^\dagger H U \right) 
		\left( U^\dagger X \right) 
		= 
		Y^\dagger \mathcal{E} Y		
	$.
	The accompanying canonical transformation $ X = U Y $ defines the eigenstates 
	$ 
		Y = 
		\begin{pmatrix}
			 \gamma		&		\gamma^\dagger	\\
		\end{pmatrix}^T 
	$
	having definite energies, stored on the diagonals of $E$ in $\mathcal{E} = \text{diag} \left( E, -E \right)$. 
	Notably, the eigenstates satisfy the fermionic anticommutation rules: $\left\{ \gamma_i, \gamma_j \right\}=0$ and $\left\{ \gamma_i, \gamma^\dagger_j \right\} = \delta_{ij}$. 
	In the Heisenberg picture they are also the eigenmodes of the time evolution operator 
	$ 
		i \partial_t  \gamma_s \left(t \right) 
		= \left[ \gamma_s \left( t \right), \hat{H} \right] 
		= E_s \gamma_s \left( t \right) 
	$
	and therefore we find their time-dependence simply from $\gamma_s \left( t \right) = \Eexp{-i E_s t} \gamma_s$, for each eigenstate $s$.
	
	While each eigenstate has a definite time-dependence, 
	the state of a condensed matter system is more complicated, consisting of mixed state that is not reducible to a single eigenstate. 
	The computation of observables therefore requires not only that the eigenvalues themselves are found,
	but the summation over the correct overlap functions of the many involved eigenstates.
	An alternative is offered by the time-dependent Green's function, as they relate directly to physical observables. 
	
	Directly corresponding to either transition or pairing amplitudes, 
	the two-point single-particle Green's functions are the 
	inhomogeneous and homogenous solutions to the equation \cite{mahan2000,Economou2006}
		$[i \partial_t - H] G(t,t') = \delta(t-t')$.
	Since this equation is linear, any linear combination of the solutions is also a solution, allowing for many possible definitions, satisfying different boundary conditions.
	We choose to work with the lesser $\mathcal{G}_{ij}^{<} (t_1,t_2)$ and greater $\mathcal{G}_{ij}^{>} (t_1,t_2)$ Green's functions as they relate directly to the the local electronic density, and as such naturally includes the quantum statistics via the Fermi-Dirac function. The same holds true for the anomalous Green's function capturing the pairing amplitudes $\mathcal{F}_{ij} (t_1,t_2)$. 
	These Green's functions are homogenous solutions, following from their definitions \cite{mahan2000,Economou2006},
		\begin{align}	
			\label{Eq:GFDEF}	
			\begin{split}	
				\mathcal{G}_{ij}^{<} (t_2,t_1) 	&\equiv i \langle c_{j}^{\dagger} (t_1) c_{i} (t_2) \rangle\,,	
				\\
				\mathcal{G}_{ij}^{>} (t_2,t_1) 	&\equiv - i \langle c_{i} (t_2) c_{j}^{\dagger} (t_1) \rangle\,,
				\\
				\mathcal{F}_{ij} (t_2,t_1) 			&\equiv i \langle c_{j} (t_1) c_{i} (t_2)  \rangle\,,
			\end{split}
	\end{align}
	and through linear combinations including appropriate factors of $i$ and the Heaviside step function $\theta(t)$ they also directly give other conventionally defined Green's functions, including the retarded, advanced, and time-ordered versions \cite{mahan2000,Economou2006}, 
	\begin{widetext}
		\begin{equation}	
			\label{Eq:GFs}	
			\begin{split}		
			 \mathcal{G}_{ij}^{R/A}	(t_2,t_1) & \equiv \pm i \theta( \pm (t_2 - t_1) )  \langle \{ c_{i} (t_2), c_{j}^{\dagger} (t_1) \} \rangle   = \pm \theta( \pm (t_2 - t_1) ) [G_{ij}^{>}(t_2,t_1) - G_{ij}^{<}(t_2,t_1)]	\,,		
				\\ 
			 \mathcal{G}_{ij}^{C} (t_2,t_1) 			& \equiv -i \langle \mathcal{T} c_{i} (t_2) c_{j}^{\dagger} (t_1)	 \rangle 	 = \theta(t_2 - t_1) G_{ij}^{>}(t_2,t_1) + \theta(t_1 - t_2 )  G_{ij}^{<}(t_2,t_1).
		\end{split}
	\end{equation}
	\end{widetext}
	Thus, from the lesser, greater, and the anomalous Green's functions, which we will calculate, we can easily find the other Green's functions. 
	
	Since the eigenstates with their definite time-dependence completely specify the system within the BdG formalism,  they also determine the time-dependence of the Green's functions. This is because, between any two generic times $t_1$ and $t_2$, a transformation to the eigenstates via  the transformation $U$ gives the explicit time-dependence through the matrix elements $v_{ij}$ and $u_{ij}$ as $c_{i} (t) = u_{is} \gamma_{s} (t) + v_{is} \gamma^{\dagger}_{s} (t) $,
	\begin{widetext}
	\begin{equation}
		\label{Eq:timeexp}	
		\begin{split}	
			 \mathcal{G}_{ij}^{<} (t_2,t_1)
			&\equiv 
			i \langle c_{j}^{\dagger} (t_1) c_{i} (t_2) \rangle
		=
			i \sum_{s}
				\left[ 		
					u_{is} u^{\dagger}_{sj}		F_\beta \left( E_s \right)  \Eexp{ - i E_s (t_2 - t_1) }
				\right.
	+
				\left.			
				v_{is} v^{\dagger}_{sj} 	F_\beta \left( -E_s \right) \Eexp{ i E_s (t_2 - t_1) }
				\right]\,,
	\\			
			 \mathcal{F}_{ij} (t_2,t_1) 
			&\equiv 
			i \langle c_{j} (t_1) c_{i} (t_2)  \rangle 
		=
			i \sum_{s}
				\left[ 			
					u_{is} v^{T}_{sj}		F_\beta \left( E_s \right) 	\Eexp{ - i E_s (t_2 - t_1) }
				\right.
		 +
				\left.		
					 v_{is} u^{T}_{sj}  	F_\beta \left( -E_s \right) \Eexp{ i E_s (t_2 - t_1) }
				\right],
		\end{split}
	\end{equation}
	\end{widetext}
	where $F_\beta \left( E_s \right) = \langle \gamma^{\dagger}_s \gamma_s\rangle $ is the Fermi-Dirac function at the inverse temperature $\beta=1/T$ and the summation runs over all eigenstates indexed by $s$.
	
	Having established the explicit expressions of the time-dependent Green's functions in Eq.~\eqref{Eq:GFDEF}, we next show that they are given by the matrix elements of  the EP $L_{\beta}(H,t)$ given by Eq.\,(\ref{Eq:EPPEXP}).
	In fact, just like the Green's functions, the EP viewed as a matrix equation is also a homogeneous solution to
	$[i \partial_t - H] L_{\beta}(H,t) = 0$, while, in addition, also the unique solution that for equal times reduces to the Fermi-Dirac projection $F_\beta \left( H \right)$ and capturing the equilibrium statistics. This fundamentally establishes a direct connection between the EP and the Green's functions given by Eqs.\,(\ref{Eq:GFDEF}). In fact, the explicit eigenbasis forms of the Green's functions, also show that the Green's functions are the  matrix elements of the EP. Because in the same eigenbasis, the Hamiltonian has a diagonal form $H = U \mathcal{E} U^\dagger$ and the matrix function $L_{\beta}(H,t) = \Eexp{ - i H t } F_\beta \left( H \right)$ is found by simply applying the same function to the real eigenvalues on the diagonal entries of $\mathcal{E}$: $L_{\beta}(H,t) = U \Eexp{ - i \mathcal{E} t } F_\beta \left( \mathcal{E} \right)  U^\dagger$. Multiplying out the matrix blocks shows that each block corresponds to one of the Green's functions when compared to the explicit expressions of Eq.~\eqref{Eq:timeexp},
	\begin{widetext}
		\begin{equation} 
			\label{Eq:LtoExp}
			\begin{split}
				L_{\beta} \left( H, t \right)
				&=
				\begin{pmatrix}
					u			&	v 		\\
					v^* 	&	u^* 	
				\end{pmatrix}
				\begin{pmatrix}
					F_{\beta}(E) \Eexp{-iEt}			&	0 		\\
					0 														&	F_{\beta}(-E) \Eexp{iEt}	
				\end{pmatrix}
				\begin{pmatrix}
					u^\dagger			&	v^T 		\\
					v^\dagger			&	u^T 	
				\end{pmatrix}
				\\
				& = 
				\begin{pmatrix}
					u F_{\beta}(E) \Eexp{-iEt} u^\dagger 	+ v F_{\beta}(-E) \Eexp{iEt} v^\dagger 			& 	u F_{\beta}(E) \Eexp{-iEt} v^T + v F_{\beta}(-E) \Eexp{iEt} u^T \\
					v^* F_{\beta}(E) \Eexp{-iEt} u^\dagger + u^* F_{\beta}(-E) \Eexp{iEt} v^\dagger 		& 	v^* F_{\beta}(E) \Eexp{-iEt} v^T + u^* F_{\beta}(-E) \Eexp{iEt} u^T
				\end{pmatrix}
			\\
			& = 
				\frac{1}{i}
				\begin{pmatrix}
					\mathcal{G}^{<} (t) 			& \mathcal{F} (t) \\
					\mathcal{F}^{\dagger} (t)	& [\mathcal{G}^{>} (t)]^{*}
				\end{pmatrix}\,.
			\end{split}
		\end{equation}	
	\end{widetext}
	Thus, the upper (lower) diagonal block in Eq.~\eqref{Eq:LtoExp} is the lesser (greater) Green's functions and the upper off-diagonal block correspond to the anomalous Green's functions with its conjugate in the lower off-diagonal block, as defined in Eq.~\eqref{Eq:GFDEF}. We therefore conclude that, by calculating the EP, we have full access to all two-point Green's functions, or equivalently, to a single-particle/hole excitation over time in a system with a time-independent Hamiltonian. We, therefore, focus on efficiently calculating the EP in the subsequent sections.

\section{Orthogonal polynomial expansion of the equilibrium propagator} \label{sec:OPandEPP}
	It is clear from  Eq.~\eqref{Eq:LtoExp} of the previous section that the time-dependent Green's functions are numerically attainable from diagonalization of the Hamiltonian. Diagonalization however incurs a $\mathcal{O}(N^3)$ computational complexity cost for a system with $N$ degrees of freedom, thus prohibiting its use for large inhomogeneous systems. Fortunately, on very general physical grounds, the full Hamiltonian matrix of any typical quantum system is highly sparse \cite{Goedecker1999} due to short-ranged hopping amplitudes and interactions. This sparseness opens up for linear-scaling iterative methods. In particular, orthogonal polynomial expansion readily produces $\mathcal{O}(N)$ methods because of the three-term recursion relationship connecting successive polynomials, while at the same time being numerically stable \cite{Goedecker1999,Bowler2012,Weise2006}. Thus, spurred by the impasse at diagonalization, we proceed with the use of orthogonal polynomials for computing the EP matrix elements.
	
	Generally, a set of orthogonal polynomials $\{ \phi_n \}$ are defined from an inner product with a weight function $w(x)$, 
	\[
		\int_{\mathcal{I}}{  \phi_{n} \left( x \right)  \phi_{m} \left( x \right)} w(x) \dd x
		= \delta_{mn} \left\|  \phi_n \right\|^2 
		,
	\]	
	on an interval $\mathcal{I}$ where they form a complete set of functions with the completeness relationship
	\begin{equation}
		\label{Eq:kernel}
			\delta \left( x - x' \right) 
			= 
			K \left( x, x' \right) 
			=  
			\sum_{n=0}^{\infty} 
			\phi_n \left( x \right) \phi_n \left( x' \right) 
			\frac{w \left( x' \right)}{\left\|  \phi_n \right\|^2 }
			\,,
	\end{equation}
	thus allowing for a generalized Fourier series expansion of generic functions on the interval.
	
	Because the BdG Hamiltonian matrix is Hermitian and assumed to have a bounded spectrum, we can employ a set of orthogonal polynomials to expand matrix functions of the BdG matrix, too. Because $H$ is in principle diagonalizable with real eigenvalues, 
	the action of the matrix function defining the EP, $L_{\beta} \left( H, t \right) = \Eexp{ - i H t } F_\beta \left( H \right)$,
	would in a diagonal representation be equivalent to applying the associated function $L_{\beta} \left( E, t \right) = \Eexp{ - i E t } F_\beta \left( E \right)$ to the real eigenvalues. Thus, to expand the EP, we can work directly with the real-domain function $L_{\beta} \left( E, t \right)$. 
	
	Notably, inserting the above polynomials completeness relationship and integrating over the kernel $K$ of Eq.~\eqref{Eq:kernel} produces the series expansion,
	\begin{equation}
		\label{Eq:LofEt}
		\begin{split}
			L_{\beta} \left( E, t \right)
			&	= 
			\sum_{n=0}^{\infty}
				\frac{\phi_n \left( E \right)}{\left\|  \phi_n \right\|^2 } 
				\int_{\mathcal{I}}
					F_\beta \left( E' \right) 
					\Eexp{ -  i E' t }
					\phi_n \left( E' \right)
					w \left( E' \right)
				\dd E'
			\\
			& =
			\sum_{n=0}^{\infty}
				\frac{\phi_n \left( E \right)}{\left\|  \phi_n \right\|^2 } 
				l^{n}_{\beta} \left( t \right)\,.
		\end{split}
	\end{equation}
	The immediate benefit of this step is that all of the dependence on time and temperature are conveniently isolated to the \textit{mode transients} defined as
	\begin{equation}
		\label{eq:lDef}
		l^{n}_{\beta} \left( t \right) =  
		\int_{\mathcal{I}}
			F_\beta \left( E \right) 
			\Eexp{ - i E t } 
			\phi_n \left( E \right)
			w \left( E \right)
		\dd E
		\,.
	\end{equation}
	The transience of $l^{n}_{\beta} \left( t \right)$ for any given $n$ is a direct result of the Riemann-Lebesgue lemma, guaranteeing that  $l^{n}_{\beta} \left( t \right) \rightarrow 0$ in the limit $t \rightarrow \infty$ \cite{Vretblad2003}. As stated, the expansion in Eq.~\eqref{Eq:LofEt} is directly carried over to the corresponding well-defined EP matrix function as
		\begin{equation} 
			\label{Eq:Lblocks}
			\begin{split}
				L_{\beta} \left( H, t \right)
				& =
				\sum_{n=0}^{\infty}
					\frac{\phi_n \left( H \right)}{\left\|  \phi_n \right\|^2 } 
					l^{n}_{\beta} \left( t \right) 
			\end{split}
		\,,
	\end{equation}
	meaning that the convenient separation of variables also extends to this orthogonal polynomial expansion of the EP. 
	
	Independently of what set of polynomials are used, the expansion in Eq.~\eqref{Eq:Lblocks} is, as shown, also a a series expansion of the Green's functions, as these are the matrix elements of the EP. The choice of the polynomials $\{ \phi_n \}$, does however affect the definitions of the functions $l^{n}_{\beta} \left( t \right)$ that depend on the set used. 
	
	To clearly show the equality between the (anomalous) Green's functions and the matrix elements of $L$, we introduce new basis vectors in the $2N$ dimensional vector space for the particle and hole part of the BdG Hamiltonian: $ [\BoldVec{e}_i]_j = \delta_{ij} $ and $ [\BoldVec{h}_i]_j = \delta_{(i+N)j} $. For with these definitions, the explicit equality is: 
		\begin{equation}
			\label{Eq:AGreen}
			\begin{split}
				\mathcal{G}_{ij}^{<} (t_2-t_1)
				& =
				i \langle
					c_j^{\dagger} \left( t_1 \right) 
					c_i \left( t_2 \right) 
				\rangle	
				= 
				i \BoldVec{e}^{\dagger}_i [ L_{\beta} \left(H,  t_2 - t_1 \right) ] \BoldVec{e}_j
				\\ & =
				i \sum_{n=0}^{\infty}
					\frac{ \BoldVec{e}^{\dagger}_i [ \phi_n \left( H \right) ] \BoldVec{e}_j  }{\left\|  \phi_n \right\|^2 } 
					l^{n}_{\beta} \left( t_2 - t_1 \right)\,,
				\\
				\mathcal{F}_{ij} (t_2-t_1) 
				&=
				i \langle
					c_j \left( t_1 \right) 
					c_i \left( t_2 \right) 
				\rangle
				= 
				i \BoldVec{e}^{\dagger}_i [ L_{\beta} \left(H,  t_2 - t_1 \right) ] \BoldVec{h}_j
				\\ & =
				i \sum_{n=0}^{\infty}
					\frac{ \BoldVec{e}^{\dagger}_i [ \phi_n \left( H \right) ] \BoldVec{h}_j  }{\left\|  \phi_n \right\|^2 } 
					l^{n}_{\beta} \left( t_2 - t_1 \right)\,.
			\end{split}
		\end{equation}
	It is this series mode expansion, which explicitly separates time and temperature effects, that is the foundation for the EPOCH method for computing time-dependent Green's functions.

\section{Legendre polynomial expansion of the equilibrium propagator} \label{sec:Legendre}
	To use the mode expansion of Eq.~\eqref{Eq:AGreen} we need to evaluate both the elements of the matrix polynomial $\phi_n \left( H \right)$ and the time-dependent mode transients $ l^{n}_{\beta} \left( t \right) $ given by Eq.~\eqref{eq:lDef}. However, attempting the direct evaluation of either one of these is both numerically unstable and inefficient. For instance, we have not found any simple closed-form analytical expressions  for $l^{n}_{\beta} \left( t \right)$. Moreover, the integrand that defines $l^{n}_{\beta} \left( t \right)$ in Eq.~\eqref{eq:lDef} becomes  for large $n$ or $t$ highly oscillatory from the energy phase factor and the high degree polynomial, thereby hindering direct numerical integration beyond the first few terms. The key to surmounting these difficulties is to use the recursive relationship between subsequent orthogonal polynomials to derive recurrence relationships for both the matrix elements $\phi_n \left( H \right)$ and the mode transients $l^{n}_{\beta} \left( t \right)$. In this way, both of these quantities are readily computed iteratively order by order, which achieves both numerical stability and efficiency in one sweep. This is a key ingredient of the EPOCH method.
	
	For definite expressions suitable for numerical treatment, we first have to choose an appropriate set of orthogonal polynomials. We choose the Legendre polynomials $P_n (x)$ defined on the interval $\mathcal{I}=[-1,1]$ having $ \left\| P_n \right\|^2 = 2/(2n+1) $. Notably, the weight function $w(x)=1$ associated with the Legendre polynomials gives equal weight and therefore also equal representational accuracy to the whole interval and spectrum. This is in contrast to the Chebychev polynomials, which add an unphysical weighting issue, see Appendix \ref{app}. Moreover, the straightforwardness of the Legendre polynomial weight function also streamlines all ensuing calculations thereby allowing us to derive key analytical expressions, as we show below. 
	
	Since any finite dimensional Hamiltonian has a bounded spectrum, we can without loss of generality assume that the spectrum is contained within the interval $\mathcal{I}$, since the Hamiltonian can always be rescaled by a constant: $H \rightarrow \tilde{H} = H/\lambda$. 
	\footnote{
		We choose to only consider a rescaling of the spectrum, because a BdG Hamiltonian always has a particle-hole symmetric spectrum. If there is no need for a BdG formalism, as for a system without superconductivity, then the spectrum can, in addition to rescaling, be translated to fit the interval $\mathcal{I}=[-1,1]$ on which the Legendre polynomials are defined. 
	}
	For a bandwidth $W$, the spectrum of $\tilde{H}$ will be entirely contained in $\mathcal{I}$ if $\lambda > W$. In what follows and for ease of presentation, we therefore assume that such a scaling has been performed and work directly with the dimensionless rescaled Hamiltonian $\tilde{H}$, which has its spectrum entirely contained in $\mathcal{I}$. Similarly, all quantities with energy dimensions are also subsequently rescaled, including the inverse temperature 
		$\beta \rightarrow \tilde{\beta} = \lambda \beta$
	and time
		$t \rightarrow \tilde{t} = \lambda t$, which all then become dimensionless quantities. 	
	Proceeding with the Legendre polynomials, below we calculate the matrix elements of $\phi_n ( \tilde{H} ) = P_n ( \tilde{H} )$ and the mode transients $ l^{n}_{\tilde{\beta}} (\tilde{t})$ in order to generate the series mode expansion in Eq.~\eqref{Eq:AGreen}. 

	\subsection{Matrix elements} \label{sec:MatEl}
		The matrix elements 
			$ 
			\BoldVec{e}^{\dagger}_i P_n ( \tilde{H} ) \BoldVec{e}_j = (P_n ( \tilde{H} ) \BoldVec{e}_i )^{\dagger} \BoldVec{e}_j
			$ 
			and 
			$
			\BoldVec{e}^{\dagger}_j P_n ( \tilde{H} ) \BoldVec{h}_j = (P_n ( \tilde{H} ) \BoldVec{e}_i )^{\dagger} \BoldVec{h}_j
			$
		that enter the mode expansions of the normal and anomalous Green's functions in Eq.~\eqref{Eq:AGreen}
		are efficiently computed by using the three-term recurrence relationship for the 
		Legendre polynomials 
		$ 
			\left( n+1 \right) P_{n+1} (x) = (2n+1) x P_n(x) - n P_{n-1}(x) 
		$.
		For this relationship connects sequential orders of the left-hand vector 
		$ \BoldVec{e}^{n}_i = P_n ( \tilde{H} ) \BoldVec{e}_i $
		via the recursion relation
		\begin{equation}
			\label{Eq:MatElRecursion}
			\left( n + 1 \right) \BoldVec{e}^{n+1}_i = (2n+1) \tilde{H} \BoldVec{e}^{n}_i - n \BoldVec{e}^{n-1}_i \,. 
		\end{equation}
		As a result, the matrix elements for all orders can be computed iteratively as the inner products 
			$\BoldVec{e}^{\dagger}_i P_n ( \tilde{H} ) \BoldVec{e}_j = (\BoldVec{e}^{n}_i)^{\dagger} \BoldVec{e}_j$
		and
			$\BoldVec{e}^{\dagger}_i P_n ( \tilde{H} ) \BoldVec{h}_j = (\BoldVec{e}^{n}_i)^{\dagger} \BoldVec{h}_j$. 
		Because these inner products are computationally cheap, the only significant computational cost 
		associated with the matrix elements is, therefore, the one sparse matrix multiplication $\tilde{H} \BoldVec{e}^{n}_i$ for each use of the recursion. In addition, once the orders of a given initial state $ \BoldVec{e}^{n}_i$ have all been computed, 
		the inner product with any end state, particle or hole, carries essentially no additional cost, making all Green's functions of that initial state available. Consequently, the complexity cost for the whole mode expansion is just $\mathcal{O}(N)$, making it very efficient.

	\subsection{Mode transients}
		The second ingredient needed for the mode expansion of Eq.~\eqref{Eq:AGreen}, in addition to the matrix elements, are the mode transients $l^{n}_{\tilde{\beta}} ( \tilde{t} )$. Their defining equation Eq.~\eqref{eq:lDef} can be seen as the Fourier transform of the product between the Legendre polynomials and Fermi-Dirac function on the interval $\mathcal{I}$, since for the Legendre polynomials the weight function is identically one. Without the Fermi-Dirac function, this Fourier transform has a simple analytical close form, because the Fourier coefficients of the Legendre polynomials are directly related to the spherical Bessel functions 
		$j_{n} ( \tilde{t} )$ 
		through 
		$
			\int_{-1}^{1}
			\Eexp{ - i \tilde{E} \tilde{t} }
			P_n ( \tilde{E} )
			\dd \tilde{E} 
			= 
			2 (-i)^n j_{n} ( \tilde{t} )
		$
		(see Ref.~\cite{GradshteynRyzhik}, Eq.~7.243.5), which is also the basis for constructing Eq.~\eqref{Eq:LegendreEvolution}.
		
		The transform of the mode transient including the Fermi-Dirac function, is however not so simple. 
		Yet, parts of the previous result can be salvaged, and the the above connection to the Bessel functions is maintained even for the non-trivial mode transients $l^{n}_{\tilde{\beta}} ( \tilde{t} )$. For because of the odd/even symmetry of the Legendre polynomials, 
		$P_n ( - \tilde{E} ) = \left( - 1 \right)^n P_n ( \tilde{E} ) $ and the partition of unity $ 1 = F_\beta ( \tilde{E} ) + F_\beta ( -\tilde{E} ) $,
		\begin{equation} 
			\label{Eq:l_conj}
			l^{n}_{\tilde{\beta}} ( \tilde{t} )
			+
			\left( -1 \right)^n
			[ l^{n}_{\tilde{\beta}} ( \tilde{t} ) ]^*
			=
			\int_{-1}^{1}
				\Eexp{ - i \tilde{E} \tilde{t} }
				P_n ( \tilde{E} )
			\dd \tilde{E}
			= 2 (-i)^n j_{n} ( \tilde{t} )
			.
		\end{equation}
		Consequently, for all inverse temperatures $\tilde{\beta}$ 		
			\begin{align*}        
					& n \quad \text{even} & \rightarrow \quad&  \mathrm{Re} \, l^{n}_{\tilde{\beta}} ( \tilde{t} ) = (-1)^{n/2} j_{n} ( \tilde{t} )\,, \\
					& n \quad \text{odd}    	&\rightarrow\quad& \mathrm{Im} \, l^{n}_{\tilde{\beta}} ( \tilde{t} ) = (-1)^{(n+1)/2} j_{n} ( \tilde{t} )\,.				
			\end{align*}			
		Thus for all $n$ either the real or imaginary part of the complex function $l^{n}_{\tilde{\beta}} ( \tilde{t} )$ are already given by the spherical Bessel functions $j_{n} ( \tilde{t} )$. 
		
		To find the remaining complementary parts of $l^{n}_{\tilde{\beta}} ( \tilde{t} )$ that is not given by the Bessel functions, we derive a new recursion relationship, by once more exploiting the recursive properties of the Legendre polynomials. Using that $ \left( 2 n + 1 \right) P_n(x) = \frac{d}{dx} \left( P_{n+1}(x) - P_{n-1}(x) \right)$ and partial integration, we find,
		\begin{equation} 
			\label{Eq:ModeRecurrence}
			\left( 2 n + 1 \right) 
			l^{n}_{\tilde{\beta}} ( \tilde{t} )
			=
			i \tilde{t}
			\left[
				l^{n+1}_{\tilde{\beta}} ( \tilde{t} )				
				-
				l^{n-1}_{\tilde{\beta}} ( \tilde{t} )
			\right]
			+
			S_{\tilde{\beta},n} (\tilde{t})
			\,.
		\end{equation}
		We recognize Eq.~\eqref{Eq:ModeRecurrence} as intimately connected to the well-known recurrence relationship for spherical ($+$) or modified spherical ($-$) Bessel functions $ z_{n+1} (x) \pm z_{n-1} (x) = (2 n + 1) z_{n} (x) /x$ (see Ref.~\cite{GradshteynRyzhik}, Eq.~8.4718.1), 
		but with an inhomogeneous source term 
		\begin{equation}		
		\label{Eq:SourceTerm}
			S_{\tilde{\beta},n} (\tilde{t})
			=
			\int_{-1}^{1}
				\left[
					P_{n+1} ( \tilde{E} )
					-
					P_{n-1} ( \tilde{E} )
				\right]
				\left(
					- F^{\prime}_{\tilde{\beta}}(\tilde{E})
				\right)
				\Eexp{ - i \tilde{E} \tilde{t} }
				\dd \tilde{E}\,.
		\end{equation}
		To further isolate the remaining part of $l^{n}_{\tilde{\beta}} ( \tilde{t} )$, we define each mode transient to be the sum of a unitary part that is simply given by the spherical Bessel function $j_{n}$ and one as of yet unknown projective part $f^{n}_{\tilde{\beta}} ( \tilde{t} )$:
		$
			l^{n}_{\tilde{\beta}} ( \tilde{t} ) = (-i)^{n} \left( j_{n} ( \tilde{t} )  + i f^{n}_{\tilde{\beta}} ( \tilde{t} ) \right)
		$.
		By this definition all of the temperature dependence is completely relegated to the projective part $f^{n}_{\tilde{\beta}} ( \tilde{t} )$.
		From Eq.~\eqref{Eq:ModeRecurrence}, it follows, as we find, that the projective part $f^{n}_{\tilde{\beta}} ( \tilde{t} )$ satisfies the inhomogeneous spherical Bessel recurrence relation:
		\begin{equation}
			\label{Eq:recurrence}
			\frac{2 n + 1}{\tilde{t}} f^{n}_{\tilde{\beta}} ( \tilde{t} )
			=
			f^{n+1}_{\tilde{\beta}} ( \tilde{t} ) + f^{n-1}_{\tilde{\beta}} ( \tilde{t} )
			+
			i^{n-1}
			\frac{ S_{\tilde{\beta},n} ( \tilde{t} ) }{\tilde{t}}
			\,.
		\end{equation}
		Notably, by exploiting the dependence of the integrand in Eq.\,(\ref{Eq:SourceTerm}) with respect to $\tilde{E}$ and $n$, we find that the inhomogeneous part in Eq.\,(\ref{Eq:recurrence}),  $i^{n-1} S_{\tilde{\beta},n} ( \tilde{t} ) / \tilde{t}$, is real for all $n$. Therefore all $f^{n}_{\tilde{\beta}} ( \tilde{t} )$ are also real.
		
		To summarize, the results of this section show that when the Legendre polynomials $P_n(x)$ are used as the orthogonal polynomial modes in the expansion of the EP in Eq.~\eqref{Eq:Lblocks}, the mode transients $l^{n}_{\tilde{\beta}}$ each decompose into a unitary and a projective transient. First, the unitary transients are given by the spherical Bessel functions $j_n(\tilde{t})$. In standard wave-packet propagation it is only this unitary part that contributes, giving the expansion presented already in Eq.~\eqref{Eq:LegendreEvolution}. Second, the projective transients $f^{n}_{\tilde{\beta}} ( \tilde{t} )$ encode the equilibrium statistics and, therefore, account for all the temperature dependence of the Green's functions. Like the unitary transients, the projective transients are real functions satisfying the spherical Bessel recursion relationship, but for $f^{n}_{\tilde{\beta}} ( \tilde{t} )$ the recursion is inhomogeneous because of the source term $S_{\tilde{\beta},n} (\tilde{t})$. With the Bessel functions $j_n(\tilde{t})$ well-known, we only have left to solve for the $f^{n}_{\tilde{\beta}} ( \tilde{t} )$ to complete our task of computing the EP and thereby all two-point Green's functions. This task requires us to evaluate the source term $S_{\tilde{\beta},n} (\tilde{t})$ in Eq.~\eqref{Eq:SourceTerm}, which we do next.

\subsection{Inhomogeneous source term} \label{sec:SourceTerm}
	At first it might not seem that the integral defining the source term $S_{\tilde{\beta},n} (\tilde{t})$ in Eq.~\eqref{Eq:SourceTerm} is any easier to evaluate than the integral originally defining the general mode transient in Eq.~\eqref{eq:lDef}. But in analogy to the Sommerfeld expansion \cite{Sommerfeld1928}, the source term $S_{\tilde{\beta},n} (\tilde{t})$ in Eq.~\eqref{Eq:SourceTerm} is readily computed as a low-temperature expansion in $1/\tilde{\beta}$, by simply exploiting that for low temperatures $F^{\prime}_{\tilde{\beta}}(\tilde{E})$ is a sharply peaked function that is exponentially localized to $\tilde{E}=0$. As a consequence, the sole contribution to the integral defining $S_{\tilde{\beta},n} (\tilde{t})$ is only coming from a small neighborhood around zero energy of width $1/\tilde{\beta}$. The energy scale associated with this width should be compared to the overall bandwidth $W$ of the system, and we can therefore extend the limits of integration to the whole real line, since for any realistic condensed matter system  $\beta W \gg 1 $.
	
	Further, because integration is a linear operation, we can additively distribute the integration over the Legendre polynomials appearing in the integral defining $S_{\tilde{\beta},n} (\tilde{t})$ in Eq.~\eqref{Eq:SourceTerm}.
	$S_{\tilde{\beta},n} (\tilde{t})$ is therefore always a linear combination of the dynamical Fermi moments $I_{\tilde{\beta}, k} (\tilde{t}) = \int_{-1}^{1} \tilde{E}^k \left(- F^{\prime}_{\tilde{\beta}}(\tilde{E}) \right) e^{ -i \tilde{E} \tilde{t}} \dd \tilde{E}$. This allows us to first consider the integral of $I_{\tilde{\beta}, k} (\tilde{t})$ and afterward take the appropriate linear combination of the result. By introducing $y=\tilde{\beta} \tilde{E}$, we find
	\begin{equation}
		\label{Eq:SommerIntegral}
		\begin{split}
			I_{\tilde{\beta}, k} (\tilde{t})		& =
			\int_{-\tilde{\beta}}^{\tilde{\beta}} 
				\left( \frac{y}{\tilde{\beta} }\right)^k
				\frac{ e^{ y - i ( y / \tilde{\beta} ) \tilde{t}} }{ \left( e^{y} + 1 \right)^{2}}
				\dd y 
			\\ & =
			\left( i \frac{\partial}{\partial \tilde{t}}\right)^k
			\int_{-\tilde{\beta}}^{\tilde{\beta}}
				\frac{ e^{ - i  ( \tilde{t}/\tilde{\beta} ) y}}{ \left( e^{y}+1 \right) \left( e^{-y} + 1 \right)} \dd y 
			\\ & \approx 
			\left( i \frac{\partial}{\partial \tilde{t}}\right)^k
			\int_{-\infty}^{\infty} 
				\frac{ e^{ - i ( \tilde{t}/\tilde{\beta} ) y}}{ \left( e^{y}+1 \right) \left( e^{-y} + 1 \right)} \dd y 
			\\ & =
			\left( i \frac{\partial}{\partial \tilde{t}}\right)^k \frac{ \pi ( \tilde{t}/\tilde{\beta} ) }{\sinh \pi ( \tilde{t}/\tilde{\beta} )} \,,
		\end{split}
	\end{equation}
	where we have first rewritten the integral as a derivative over the integral;	next, because the integrand is exponentially suppressed in both directions, we have extended the integration to the whole real line allowing us then to perform the contour integration in the last step. In restored units, the contributions from outside the original interval are exponentially suppressed in the ratio of the bandwidth $W$ to the temperature by a factor of $e^{-\beta W}$. Thus, the approximation introduced by extending the integral is inconsequential for any realistic condensed matter system where $\beta W \gg 1 $. For very large moment ordinals $k$, however, the integrand eventually grows outside even this interval and the assumption in Eq.~\eqref{Eq:SommerIntegral} breaks down for very large $k \gtrsim e \beta W$. Still, using the expression of Eq.~\eqref{Eq:SommerIntegral} for the Fermi moments produces an asymptotic expansion that is very accurate for all moments with $k \lesssim \beta W$.
	
	The source term $S_{\tilde{\beta},n} (\tilde{t})$ is now a linear combination of the dynamical Fermi moments $I_{\tilde{\beta}, k} (\tilde{t})$ of  Eq.~\eqref{Eq:SommerIntegral}, where the coefficients are given by the explicit representation of the Legendre polynomials, $P_n(x) = 2^n \sum_{k=0}^n \binom{n}{k} \binom{\frac{n+k-1}{2}}{n} x^k$. Thus, we finally arrive at a low-temperature asymptotic expansion of the source term,
	\begin{equation}
		\label{Eq:SofBetaT}
			S_{\tilde{\beta},n} (\tilde{t}) 
			=
			\sum_{k=0}^{n+1}
			\frac{
				(2 n + 1) 2^n \Gamma \left( \frac{k+n}{2} \right) 
			}{
				k! \Gamma \left( \frac{k-n}{2} \right) \Gamma \left( 2 + n - k \right)
			}
			I_{\tilde{\beta}, k} (\tilde{t})\,,
	\end{equation}
	where we use the gamma function $\Gamma$. In the zero temperature limit, $\tilde{\beta} \rightarrow \infty$, it is clear that all the dynamical Fermi moments become time-independent and only the first term is non-zero, $I_{\tilde{\beta}, k} (\tilde{t}) \rightarrow \delta_{k 0}$. In this case therefore $S_{\infty,n} (\tilde{t})$ is also time-independent and non-zero only for odd $n$, and reduced simply to: 
		$
		S_{\infty,n} = 
		(
			2^n (2 n + 1) \Gamma \left( \frac{n}{2} \right) 
		)/(
			\Gamma \left( - \frac{n}{2} \right) \Gamma \left( 2 + n \right)
		)
	$. 
	
	To summarize, because the derivative of the Fermi function $F^{\prime}_{\tilde{\beta}}(\tilde{E})$ is sharply peaked around zero for all temperatures that are small compared to the bandwidth $W$, we can compute the inhomogeneous source terms $S_{\tilde{\beta},n} (\tilde{t})$ as an asymptotic series expansion in the dynamical Fermi moments $I_{\tilde{\beta}, k} (\tilde{t})$, which is rapidly converging.
As a consequence, we can either use simply the zero temperature source term $S_{\infty,n}$ or the first terms of the low temperature expansion in Eq.~\eqref{Eq:SofBetaT} to always find a good numerical approximation $S_{\tilde{\beta},n} (\tilde{t})$ for realistic temperatures.
	
\subsection{Calculating mode transients} \label{sec:NumericalProjective}
	With the inhomogeneous source term $S_{\tilde{\beta},n} (\tilde{t})$ determined in the previous subsection, we find the total mode transients $l^{n}_{\tilde{\beta}} ( \tilde{t} )$ by solving the recurrence relation Eq.~\eqref{Eq:recurrence} for the projective transients $f^{n}_{\tilde{\beta}} ( \tilde{t} )$. However, when attempting to numerically solve this recurrence relationship, care must  be taken. 
	
	In fact, if one attempts to solve it by simple forward propagation, calculating the next term from the previous two, then rounding errors inevitably and disastrously accumulate with each step. This is a well-known danger of second order recurrence relationships, including that of the Bessel functions \cite{Gautschi1967}. For our purposes, the instability of Eq.~\eqref{Eq:recurrence} is most readily understood starting from its underlying homogeneous Bessel recurrence relationship with its two linearly independent solutions, commonly called $J_n$ and $Y_n$. Here, the massive domination of one solution over the other ($ \left| Y_n (x) / J_n (x) \right| \sim  2 (n!)^2 / (x/2)^{2n} \rightarrow \infty$, as $n$ goes to infinity) implies that even if just a small rounding error falls outside the subspace of $J_n$ during numerical stepping procedure, then this spillover will quickly grow and overtake the calculated solution such that any resemblance to the desired $J_n$ is disastrously lost. 
	
	Fortunately, there already exists a stable algorithm for solving linear recurrence relationships, including inhomogeneous relationships. Central to this algorithm is the reformulation of the recurrence relationship as an auxiliary linear boundary value problem; the left boundary is set by the true initial-value of $f^0$, but the right boundary, at a sufficiently large $M$, is instead set to an artificially imposed value of $f^M = 0$. Solving this boundary value problem has been shown to converge to the minimal solution, as the right bound is moved further away with larger $M$ \cite{olver1967numerical, Oliver1968}. The recurrence equation of Eq.~\eqref{Eq:recurrence} for the projective transients $f^{n}_{\tilde{\beta}} ( \tilde{t} )$ is therefore easily solved as a boundary value problem, given that we provide the initial condition $f^{0}_{\tilde{\beta}} ( \tilde{t} ) = \int_{-1}^{1} F_{\tilde{\beta}} (\tilde{E}) \text{sin} (-\tilde{E} \tilde{t}) \dd \tilde{E}$ by direct integration.
	
	\begin{figure}[b]
		\centering
		\includegraphics[width=0.45\textwidth]{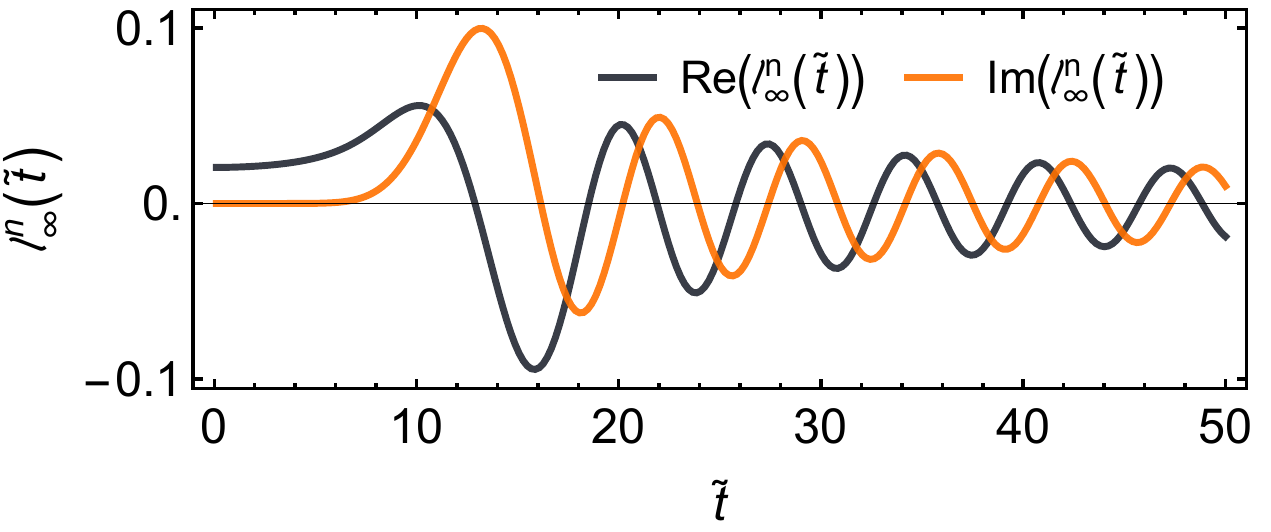}
		\caption{%
			Plot of the real part, $f^{n=11}_{\infty} ( \tilde{t} )$, and imaginary part, $j_{n=11} (\tilde{t})$, of the 
			mode transient $l^{n=11}_{\tilde{\beta}} ( \tilde{t} )$ at zero temperature, $\tilde{\beta}=\infty$.
		}
		\label{fig:fig1}
	\end{figure}

	To illustrate the mode transients we plot in Fig.~\ref{fig:fig1} the real and imaginary parts of $l^n_{\tilde{\beta}=\infty}(\tilde{t})$ for $n = 11$, where the real part is the projective transient $f^{n}_{\tilde{\beta}} ( \tilde{t} )$ generated by Eq.~\eqref{Eq:recurrence}, while the imaginary part is the unitary transient, the spherical Bessel function $j_{n}(\tilde{t})$.  As seen, the projective transient resembles a phase shifted version of its companion unitary transient, characterized by an oscillating tail, decaying with time $\sim 1 / \sqrt{\tilde{t}} $ for large $\tilde{t}$. Similarly, the envelopes of both $f^{n}_{\tilde{\beta}} ( \tilde{t} )$ and $j_n (\tilde{t})$ are of maximal height in the neighborhood of $\tilde{t} \sim n$. Thus for a given $\tilde{t}$, it is precisely the orders with $n \sim \tilde{t}$ that mainly contributes to the time dependence in the series expansion of the EP. 
	
	Generally therefore, it is the envelopes of the two transient functions that dictate which terms of the expansion make a significant contribution to the mode expansion and in turn also sets the accuracy of the Green's functions when computed from the truncated mode expansion. In this regard, there is a distinction between the unitary and projective transients. For while $j_n (\tilde{t})$ is exponentially suppressed for $\tilde{t} \lesssim n$, the $f^{n}_{\tilde{\beta}} \left( 0 \right)$ can take on finite initial values, meaning that even for short evolution times a finite number of terms, to be specified in Sec.~\ref{sec:Truncation}, are needed to capture the projective part. 

\subsection{Static limit} \label{sec:StaticLimit}
	While our main focus in this work is on the dynamical properties and the time-domain Green's functions, the static limit $\tilde{t} \rightarrow 0$ is also of interest for two reasons. First, when $\tilde{t} \rightarrow 0$, the EP reduces to the density matrix $L_{\tilde{\beta}}(\tilde{H}, 0) = F_{\tilde{\beta}}(\tilde{H})$ and its elements are the $\tilde{t} = 0$ Green's functions, measuring the densities of the system: 
		the local electron density and equal-time pair amplitudes, respectively. To have direct access to these with in the same formalism is of course of great utility. Secondly, as seen in the previous subsection, the finite initial values of $f^{n}_{\tilde{\beta}} \left( 0 \right)$ inform the rate of convergence of the mode expansion; because, unlike the case with wave-packet propagation, even at $\tilde{t}=0$ the Fermi-Dirac statistics is encoded in $f^{n}_{\tilde{\beta}} \left( 0 \right)$. 
		
	Conveniently, in the $\tilde{t} \rightarrow 0$ limit the mode transient recursion Eq.~\eqref{Eq:ModeRecurrence}  decouples and consequently the mode transients are proportional to the source terms:
		$ l_{\tilde{\beta}}^{n} (0) = i^{n-1} f^{n}_{\tilde{\beta}} \left( 0 \right) =  S_{\tilde{\beta},n} (0) / (2n+1) $ 
	for all $n \geq 1$, whence $j_n (0)=0$. 
	In addition, the static limit of the dynamical Fermi moments of Eq.~\eqref{Eq:SommerIntegral} is 
	$ \lim_{\tilde{t} \to 0} I_{\tilde{\beta}, k} (\tilde{t}) = ( 2 \pi / i \tilde{\beta} )^k (2^{1-k}-1) B_{k}$, 
		where $B_{k}$ are the Bernoulli numbers (follows from the Taylor expansion
		$
			x/\sinh(x) = 
			1 + \sum_{n=1}^\infty 2 (1-2^{2n-1}) B_{2n} x^{2n} / (2n)!
		$).
	Inserting these static Fermi moments in Eq.~\eqref{Eq:SofBetaT} to get the source terms $S_{\tilde{\beta},n} (0)$, the static limit of the mode expansion of EP in Eq.~\eqref{Eq:Lblocks} gives a polynomial expansion of the density matrix:
	\begin{equation}	
		\label{Eq:TDMAT}
		\begin{split}
			 L_{\tilde{\beta}}(\tilde{H}, 0)
			&= F_{\tilde{\beta}}(\tilde{H}) 
			\\ & =
			\frac{P_{0}(\tilde{H})}{\left\| P_0 \right\|^2}
			+
			\sum_{n=1}^{\infty} 
			\frac{P_{n}(\tilde{H})}{\left\| P_n \right\|^2}			
			\times
			\\ & \quad 
			\sum_{k=0}^{n+1}
				\frac{
					2^{n} \Gamma \left(\frac{n+k}{2}\right) \left(2^{1-k}-1\right) B_{k}
				}{
					\mathrm{k!} \Gamma \left(\frac{k-n}{2}\right) \Gamma(2+n-k)
				} 
				\left( \frac{2 \pi}{i \tilde{\beta}} \right)^{k} \,,
		\end{split}
	\end{equation}
	where $l^{0}_{\tilde{\beta}} (0) = 1$ is the origin of the $n=0$ term. In the zero temperature limit, $T=0$, only the first $k=0$ term remains
	\begin{equation}	
		\label{Eq:ZDMAT}
			\begin{split}
				 L_{\infty}(\tilde{H}, 0)
				&= F_{\infty}(\tilde{H}) 
				\\ & =
				\frac{ P_{0}(\tilde{H}) }{ \left\| P_0 \right\|^2 }
				+
				\sum_{n=1}^{\infty}
				\frac{ P_{n}(\tilde{H}) }{\left\| P_n \right\|^2}	
				\frac{ 
						2^{n} \Gamma \left( \frac{n}{2} \right)
					}{
						\mathrm{k!} \Gamma \left( \frac{-n}{2}\right) \Gamma(2+n)
				}\,.
			\end{split}
	\end{equation}
	Both Eqs.~\eqref{Eq:TDMAT} and \eqref{Eq:ZDMAT} are particularly useful expansions of the density matrix (Fermi operator) even at finite temperatures compared to the pioneering expansions of Ref.~\cite{PhysRevLett.73.122,PhysRevB.51.9455}. Thus from Eqs.~\eqref{Eq:TDMAT} or \eqref{Eq:ZDMAT} the density expectation values of the system are given by  $ \langle c_i^{\dagger} c_j \rangle = \BoldVec{e}^{\dagger}_i [ L_{\tilde{\beta}} (\tilde{H},0 ) ] \BoldVec{e}_j	$ and $ \langle	c_i	c_j \rangle = \BoldVec{e}^{\dagger}_i [ L_{\tilde{\beta}} (\tilde{H},0 ) ] \BoldVec{h}_j	$. 
		Because the assumption underlying the low-temperature expansion of the Fermi moments in Eq.~\eqref{Eq:SommerIntegral}
		eventually breaks down for very large moment ordinals $k$, the absolute values of the static moments $I_{\tilde{\beta}, k} (0)$ pass through a minimum at $k \sim \beta W$ before eventually growing again. In turn, the finite temperature series of Eq.~\eqref{Eq:TDMAT} is not formally convergent to all orders. 
		Still, because the minimum of $I_{\tilde{\beta}, k} (0)$ is exponentially suppressed in the ratio of the bandwidth to the temperature as $e^{-\beta W}$,  the series of Eq.~\eqref{Eq:TDMAT} is an exponentially accurate asymptotic expansion for the density matrix. In the zero temperature limit, the asymptotic limit of the coefficients in Eq.~\eqref{Eq:ZDMAT} decay as $n^{-3/2}$, demonstrating the norm convergence of this series.

\section{Truncation of equilibrium propagator expansion} \label{sec:Truncation}
	\begin{figure*}
		\centering
			\includegraphics[width=1.00\textwidth]{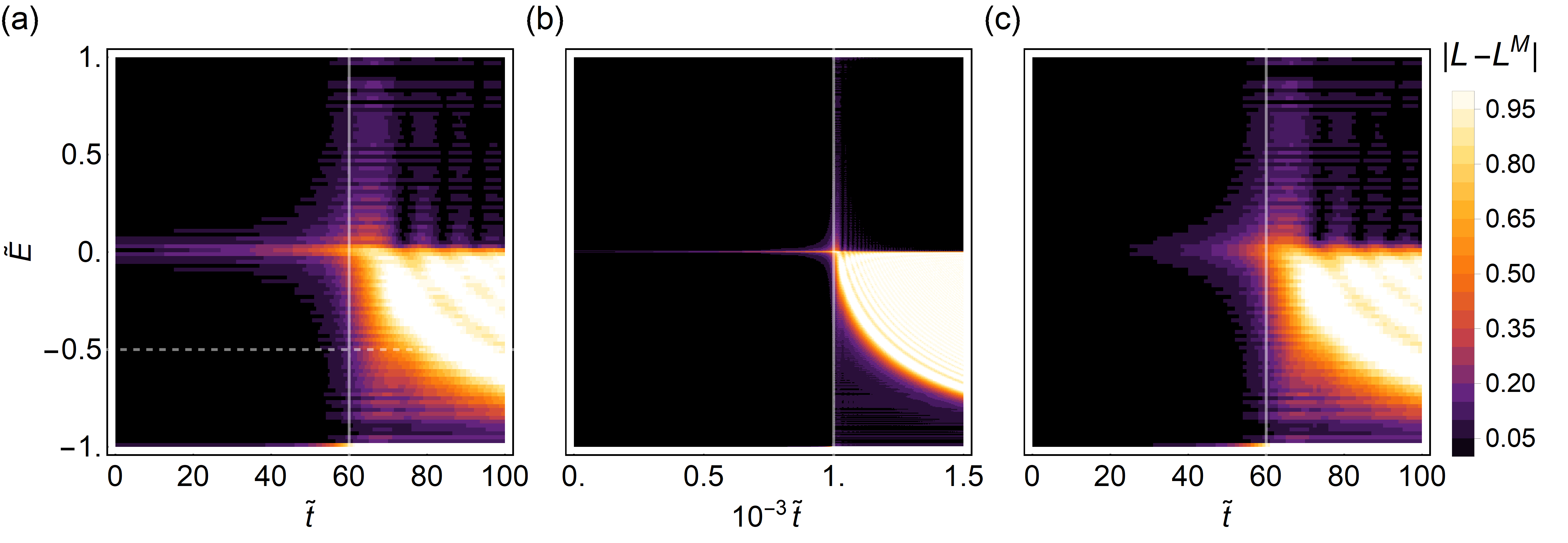}
			\caption{%
				Finite moment truncation error $\left| L_{\infty} (\tilde{E},\tilde{t}) - L_{\infty}^{M} (\tilde{E},\tilde{t}) \right|$ of the zero temperature mode expansion $L_{\infty} (\tilde{E},\tilde{t})$ as a function of time $\tilde{t}$ and energy $\tilde{E}$ for (a) $M=60$ moments and (b) $M=1000$ moments. (c) Finite temperature $\tilde{\beta}=60$ truncation error $\left| L_{\tilde{\beta}} (\tilde{E},\tilde{t}) - L_{\tilde{\beta}}^{M} (\tilde{E},\tilde{t}) \right|$ for $M=60$ moments. In each figure, the time $\tilde{t}=M$ is highlighted by a vertical white line, while dashed white line in (a) indicates the energy cut used in Fig.~\ref{fig:fig3}(c). 
			}
			\label{fig:fig2}
	\end{figure*}

	In the previous sections we have derived an exact expansion of the time-dependent two-point Green's functions using Legendre polynomials. While this series is, in principle, exact, in practice the expansion of the EP $L_{\tilde{\beta}} ( \tilde{E}, \tilde{t} )$ in Eq.~\eqref{Eq:LofEt} has to be truncated at a finite number of terms in order to produce a numerically useful algorithm. 	
	Because the unitary part $j_n (\tilde{t})$ only contributes significantly after $\tilde{t} \gtrsim n$, 
	the truncation of this series to order $M$ yields a remarkably accurate approximation of the time-dependence
	for all $\tilde{t} \lesssim M$. We illustrate this in Fig.~\ref{fig:fig2}(a,b) where we plot the numerical error $\left| L_{\infty} (\tilde{E},\tilde{t}) - L_{\infty}^{M} (\tilde{E},\tilde{t}) \right|$ for $M = 60$ and $M = 1000$, respectively.  
	The only exception is a narrow energy window $\delta \tilde{E} \sim 1/M$ closest to the Fermi surface at $\tilde{E}=0$, where the representation of the step-like Fermi-Dirac function in the projective transients $f^{n}_{\tilde{\beta}} ( \tilde{t} )$ starts to deviate because of Gibbs phenomenon, see below in Sec.~\ref{sec:Gibbs}. This is similar to the residual norm of the truncated static mode expansion, which also vanishes as  
	$
		\left\| L_{\infty} (\tilde{E},0) - L_{\infty}^{M} (\tilde{E},0) \right\|  \sim 1/M 
	$, because the coefficients of Eq.~\eqref{Eq:ZDMAT} vanish as $n^{-3/2}$.	
	However, this error source at low energies is inconsequential for any gapped system, such as insulators or superconductors with an energy gap $\Delta$. Simply put, since there are no contributing states in the misrepresented region around the Fermi level, there can be no numerical distortion of the Green's functions. For example, if  for a superconductor we assume a temperature below the transition temperature, we are guaranteed that the time-dependent Green's functions computed using the truncated mode expansion of Eq.~\eqref{Eq:AGreen} have excellent fidelity for all $\tilde{t} \lesssim M$, as long as $\Delta / W \lesssim 1/M$, where $W$ is the bandwidth.

\subsection{Gibbs phenomenon} \label{sec:Gibbs}
The numerical errors found at the Fermi level and at low temperatures in Fig.~\eqref{fig:fig2} can be understood as a manifestation of the Gibbs phenomenon; general ringing or oscillations found when an orthogonal series expansion is used to capture a step function. 		
	In short, therefore, the Gibbs phenomenon appears near the sharp Fermi surface in the finite series representation. 
	As such, it is however only relevant for systems with a finite density of states near the Fermi energy. 
	Thus, the Gibbs phenomenon is not relevant for an insulating or gapped superconducting system, but it could matter for a metallic systems.
	
\begin{figure}
		\centering
			\includegraphics[width=0.45\textwidth]{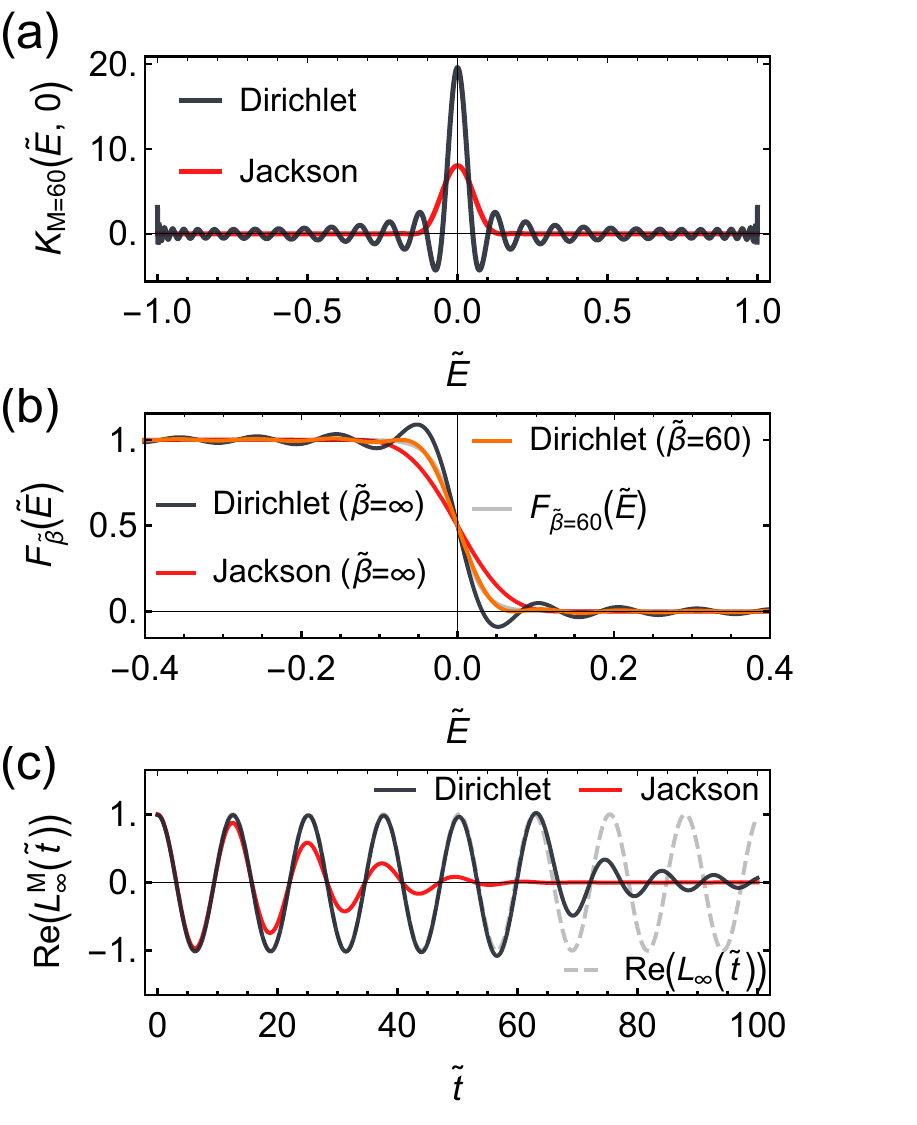}
		\caption{%
			(a) Plot of the Dirichlet $K_M^D (0,\tilde{E})$ and Jackson Kernel $K_M^J (0,\tilde{E})$ using $M=60$ moments.  
			(b) Polynomial approximation with $M=60$ moments of the Fermi-Dirac distribution $ F_{\tilde{\beta}}(\tilde{E}) = (e^{\tilde{\beta} \tilde{E}} + 1 )^{-1}$ at zero temperature $\tilde{\beta}=\infty$ (c.f.~Eq.\eqref{Eq:ZDMAT}) for both the Dirichlet or the Jackson Kernel as well as the finite temperature approximation $(\tilde{\beta}=60)$ using the Dirichlet kernel. For comparison the true $F_{\tilde{\beta}=60}(\tilde{E})$ is also show (gray line).
			(c) Real part of the finite moment approximation $L_{\infty}^{M=60} (\tilde{E}=0.5, \tilde{t} )$ using the Dirichlet and Jackson Kernels compared to the true function $L_{\infty}(\tilde{E}=0.5,\tilde{t})$. 
		}
		\label{fig:fig3}
	\end{figure}

	The Gibbs phenomenon comes about because truncating $L_{\tilde{\beta}} ( \tilde{E}, \tilde{t} )$ in Eq.~\eqref{Eq:LofEt} at the finite order $M$,
	amounts to replacing the underlying true integration kernel of Eq.~\eqref{Eq:kernel} with the truncated Dirichlet kernel,
	$
		K_M^D (x,x') 
		= 
		\sum_{n=0}^{M} 
		P_n \left( x \right) P_n \left( x' \right) / {\left\|  P_n \right\|^2 }
	$.
	This replacement is a controlled approximation that does narrow in on the Dirac delta function with a diminishing width of $\delta \tilde{E} \sim 1/M$ \cite{Kunishima2002}, as illustrated by the black curve in Fig.~\ref{fig:fig3}(a), where we plot $K_M^D (0,\tilde{E})$ for $M=60$. Nonetheless, it is also apparent that this kernel is subject to Gibbs oscillations. A direct consequence of these oscillations are the overshooting ripples in the finite series Dirichlet kernel approximation of the discontinuous $T=0$ Fermi-Dirac function seen in Fig.~\ref{fig:fig3}(b). 
	
	Echoing standard Fourier theory \cite{Vretblad2003}, the standard approach for combating the Gibbs phenomenon is to replace the naive Dirichlet kernel with a suitable summability kernel \cite{Weise2006}. For instance, the Jackson kernel is a common choice, which is an everywhere positive optimal kernel with a minimal squared peak width, as displayed by the red curve in Fig.~\ref{fig:fig3}(a). 
	It is defined by reweighing the terms of the series	
	$
		K_M^J (x,x') = \sum_{n=0}^{M} g_n^J P_n(x) P_n(x') / \left\| P_n \right\|^2
	$ 
	with the diminishing weights  
	$
		g_n^J 
		= 
		(	
			(M+1-n) \cos \left( \frac{ \pi  n }{ M + 1 } \right) 
			+
			\cot \left(\frac{\pi }{M+1}\right) \sin \left(\frac{\pi  n}{M+1}\right) 
		) 
		/ 
		( M + 1 )
	$.
	In Fig.~\ref{fig:fig3}(b) the red curve shows how adopting the Jackson kernel results in an oscillation-free approximation of the $T=0$ Fermi-Dirac distribution function. While this is a satisfactory solution for the static representation of the Fermi-Dirac function, the diminished weight given to the high-order terms by the Jackson kernel (and also by other traditional summability kernels such as the Fej\'er Kernel or the Lorentz kernel \cite{Weise2006}) simultaneously means that the convergence of the representation is severely damaged when extended to time-evolution. For when evolved to a time $\tilde{t}$, then it is the terms of order $n \sim \tilde{t}$ that are the largest and have the most significant contribution to the representation. Thus, the diminished weight given to the higher order terms by the summability kernels, causes the approximation of $L_{\tilde{\beta}} ( \tilde{E}, \tilde{t} )$ to deviate even after just a short time-evolution, as we display in Fig.~\ref{fig:fig3}(c) along with this effect. In contrast, the Dirichlet kernel is, also in Fig.~\ref{fig:fig3}(c), seen to accurately represent $L_{\tilde{\beta}} ( \tilde{E}, \tilde{t} )$ for all $\tilde{t} \lesssim n$, while the Jackson kernel fails even for small $\tilde{t}$. Thus, the standard approach for circumventing the Gibbs phenomenon is not useful when going beyond the static Fermi-Dirac function to the time-evolution.

	Although we find that the widely accepted technique of kernel resummation is not appropriate for time-evolution, 
	we nonetheless find a working solution to the Gibbs phenomenon by transitioning to finite temperatures.
	At non-zero temperature, the step discontinuity of the Fermi-Dirac distribution is replaced by a smooth transition of width $\delta \tilde{E} \sim \tilde{\beta}$. The distribution therefore becomes amenable to an accurate representation by a polynomial approximation of order $M \sim \tilde{\beta}$, as in Eq.~\eqref{Eq:TDMAT}. We illustrate this in Fig.~\ref{fig:fig3}(b) where the $\tilde{\beta}=60$ distribution is plotted with $M=60$ moments. 
	
	Importantly, by constructing the finite temperature representation of the EP by solving the recurrence relationships of Eq.~\eqref{Eq:recurrence} for the projective mode transients with the first few terms of the low-temperature expansion Eq.~\eqref{Eq:SofBetaT},
	we avoid the problems arising from Gibbs phenomenon, and the resulting representation of the EP has all the accuracy of its zero-temperature version at all times $\tilde{t} \lesssim n$ away from the Fermi surface.	But in addition, the representation is now also accurate around the step of the Fermi-Dirac distribution. Meaning that, $L_{\tilde{\beta}} ( \tilde{E}, \tilde{t} )$ is accurate for all $\tilde{t} \lesssim n/2$, including the energies near the Fermi surface, as is illustrated in Fig.~\ref{fig:fig2}(c). As a consequence, even quantitative calculations in systems with a finite density of states near the Fermi energy are possible at any finite temperature.

\section{Summary of EPOCH method} \label{sec:algorithm}	
	In summary, we have shown that time-domain Green's functions of a time-independent Hamiltonian $H$ are efficiently and accurately computed as the matrix elements of the EP by the orthogonal polynomial expansion Eq.~\eqref{Eq:Lblocks} using Legendre polynomials. We call this method EPOCH, standing for Equilibrium Propagator by Orthogonal polynomial CHain. For additional clarity, we here summarize the steps of the EPOCH method:
	\begin{enumerate}
		\item%
				Rescale the Hamiltonian through a change of energy units $H \rightarrow \tilde{H} = H/\lambda$ so that the spectrum of $\tilde{H}$ is entirely contained in the interval $[-1,1]$ where the Legendre polynomials are defined. For a bandwidth $W$, this is guaranteed if $\lambda > W$.
		\item%
			For all inverse-temperatures $\lambda \beta$ and time steps $\lambda t$ for which the Green's functions are to be calculated, solve the recurrence equation Eq.~\eqref{Eq:recurrence} for the projective transients $f^{n}_{\lambda  \beta} \left( \lambda t \right)$ as a boundary value problem with the method of Ref.~\cite{olver1967numerical, Oliver1968}. For the source term $S_{\lambda \beta,n}$ use either the zero temperature source terms $S_{\infty,n}$ or the first terms of the low-temperature expansion Eq.~\eqref{Eq:SofBetaT}.
		\item%
			Construct the total mode transient as the sum of the unitary and projective transients: $l^{n}_{\lambda \beta} ( \lambda t) = (-i)^{n} ( j_{n} (  \lambda t)  + i f^{n}_{\lambda \beta} \left( \lambda t \right) )$, where the spherical Bessel functions $j_n( \lambda t)$ are already known.
		\item%
			For all Green's functions $\mathcal{G}^{<}_{ij}$ or $\mathcal{F}_{ij}$ of interest compute the matrix elements 
			$\BoldVec{e}^{\dagger}_i P_n ( \tilde{H} ) \BoldVec{e}_j = (\BoldVec{e}^{n}_i)^{\dagger} \BoldVec{e}_j$ 
			and                                       
			$\BoldVec{e}^{\dagger}_i P_n ( \tilde{H} ) \BoldVec{h}_j = (\BoldVec{e}^{n}_i)^{\dagger} \BoldVec{h}_j$, respectively,
			using the recursion relationship in Eq.~\eqref{Eq:MatElRecursion}) for the vector 
			$ \BoldVec{e}^{n}_i = P_{n} \left( H \right) \BoldVec{e}_i$, starting from the particle vector of the injection site $ \BoldVec{e}^{n=0}_i = \BoldVec{e}_i $.	
		\item%
			Construct the normal and anomalous Green's functions in the time-domain with $t=t_2-t_1$ by the mode expansion of Eq.~\eqref{Eq:AGreen}:
			\indent \begin{equation}			
			\label{Eq:AlgoGandF}
				\begin{split}
					\mathcal{G}^{<}_{ij} (t) 
						= i \langle c_{j}^{\dagger} (t_1) c_{i} (t_2) \rangle	
						\approx	i \sum_{n=0}^{M} [ (\BoldVec{e}^{n}_i)^{\dagger} \BoldVec{e}_j ] l^{n}_{\lambda \beta} (\lambda t)\,,
					\\
						\mathcal{F}_{ij} (t) 
						= i \langle c_{j} (t_1) c_{i} (t_2)  \rangle
						\approx	i \sum_{n=0}^{M} [ (\BoldVec{e}^{n}_i)^{\dagger} \BoldVec{h}_j ] l^{n}_{\lambda \beta} (\lambda t)\,.
				\end{split}
			\end{equation}
	\end{enumerate}
	
	The mode expansion in Eq.~\eqref{Eq:AlgoGandF} using $M$ modes is able to faithfully represent the Green's functions for $\lambda t \lesssim M$, except in a narrow energy interval of width $\lambda / M$ around the Fermi energy. If the system has an energy gap $\Delta E$, then the  Green's functions are faithfully represented for all $\lambda t \lesssim M$ as long as $ \Delta E \lesssim \lambda / M$. Even for systems with a finite density of states near the Fermi energy, it is possible to faithfully represent the finite temperature Green's functions down to temperatures of $\lambda \beta \gtrsim M$ and out to $\lambda t \lesssim M/2$ in time, as illustrated in Fig.~\ref{fig:fig1}.
	
		\begin{figure*}[htb]
		\centering
			\includegraphics[width=0.85\textwidth]{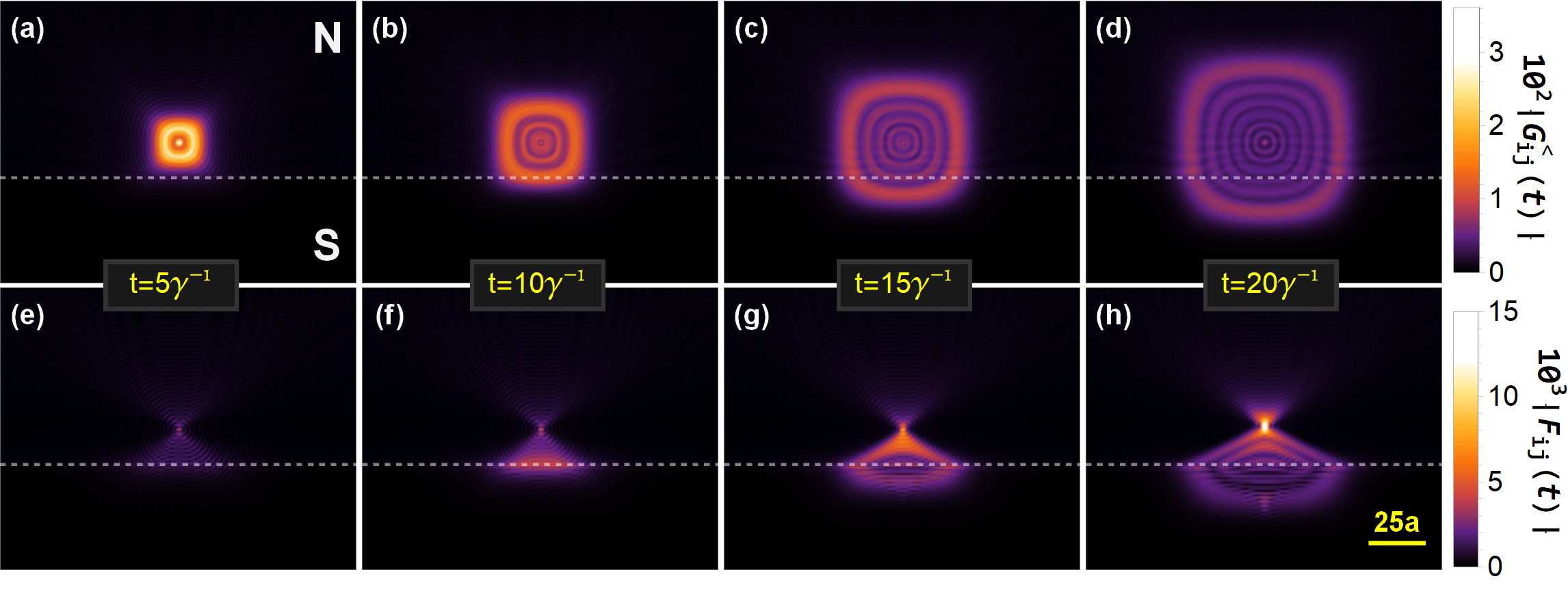}
		\caption{%
			Snapshots in time of the evolution of both the particle-particle $|\mathcal{G}^{<}_{ij}(t)|$ (a-d) and the particle-hole $|\mathcal{F}_{ij}(t)|$ (e-h) correlation amplitudes for a particle inserted on the normal side near the interface of a 2D NS junction, with interface indicated by dashed white line. The square lattice is of dimensions $L^2=(201a)^2$ and extends beyond the regions shown and the spatial scale in indicated by bar in (h). The evolution was computed using the EPOCH method at zero temperature with $M=1000$ moments applied to the Hamiltonian in Eq.~\eqref{Eq:ModelH} at the chemical potential $\mu = 2 \gamma$ and pairing potential $\Delta=0.25 \gamma$ inside the superconductor.
		}
		\label{fig:M2D}
	\end{figure*}
		
	The computational cost of the EPOCH method is mainly set by the computation of the matrix elements of $P_n ( \tilde{H} )$, where each additional moment requires a matrix-vector multiplication by $\tilde{H}$. For a sparse matrix, this is an $\mathcal{O} (N)$ operation, that furthermore need only be done for the sites of interest and is also easily parallelized over multiple CPU or GPU cores. Moreover, the EPOCH methods also scales linearly in propagated time, i.e., running the time evolution for twice as long will, all else being equal, take twice as long to compute. Thus, even for systems with large dimensions $N$ the time-domain Green's function for large relative time differences are efficiently and easily computed computed with the EPOCH method.
			
\section{Example: Superconductor-normal metal interface}
	\label{sec:SN}	
To illustrate the vast possibilities of the EPOCH method, while still studying a well known, but large system, we calculate the time-dependence of the normal and anomalous Green's functions in a superconductor-normal metal (SN) junction in both 2D and 3D. At the SN interface, an incident particle can, in addition to normal transmission and reflectance, also exhibit Andreev reflection \cite{andreev1964thermal}, where a particle is converted to a hole and a Cooper pair is transferred to the superconductor. Andreev reflection is thus the process responsible for the superconducting proximity effect, which not only changes the properties of SN junctions but also directly leads to the finite Josephson current in junctions with two superconductors leads.
 Andreev refection is thus experimentally detectable in tunneling experiments \cite{Rowell1973}, and with dual contact measurements the Andreev reflections are even by themselves directly observable \cite{Benistant1983,Benistant1985}. 
 Scattering at SN interfaces is therefore an interesting fundamental and experimentally measurable process, for which EPOCH is ideally suited. Specifically, we consider a finite sized tight-binding model on the square and cubic lattice, respectively, described by
	\begin{equation}	
		\label{Eq:ModelH}
		\begin{split}		
			\hat{H} 
			&= 
				-\gamma \sum_{\langle i, j \rangle, \sigma} [ c^{\dagger}_{i \sigma} c_{j \sigma} + \text{H.c.}]
				-\mu \sum_{i, \sigma} c^{\dagger}_{i \sigma} c_{i \sigma}
			\\ & 
				+ \Delta \sum_{i} \theta ( -\hat{x} \cdot \vec{r}_i ) [  c_{i \uparrow} c_{i \downarrow}  + \text{H.c.}]\,,
		\end{split}
	\end{equation}
	where the particles (electrons) are created by the operator $c^{\dagger}_{i \sigma}$ at the lattice site $i$ with spin $\sigma =\uparrow,\downarrow$ at the position $\vec{r}_i = a (i_x \hat{x} + i_y \hat{y})$ in 2D and $\vec{r}_i = a (i_x \hat{x} + i_y \hat{y} + i_z \hat{z})$ in 3D, where $a$ is the lattice constant and the total length of each sides of the lattice is $L$. With respect to the BdG form of Eq.~\eqref{Eq:BdGForm}, the diagonal normal part $\hat{H}_0$ includes the first two terms of $\hat{H}$ with the nearest-neighbor hopping amplitude $\gamma$ and the chemical potential $\mu$. Similarly, the last term is the off-diagonal part $\hat{\Delta}$ of the BdG form, representing a constant superconducting order parameter of finite amplitude $\Delta$ in half of the system $x<0$, given by the Heaviside step function. This generates an NS interface at $x=0$.	
	\begin{figure*}
		\centering
			\includegraphics[width=1.00\textwidth]{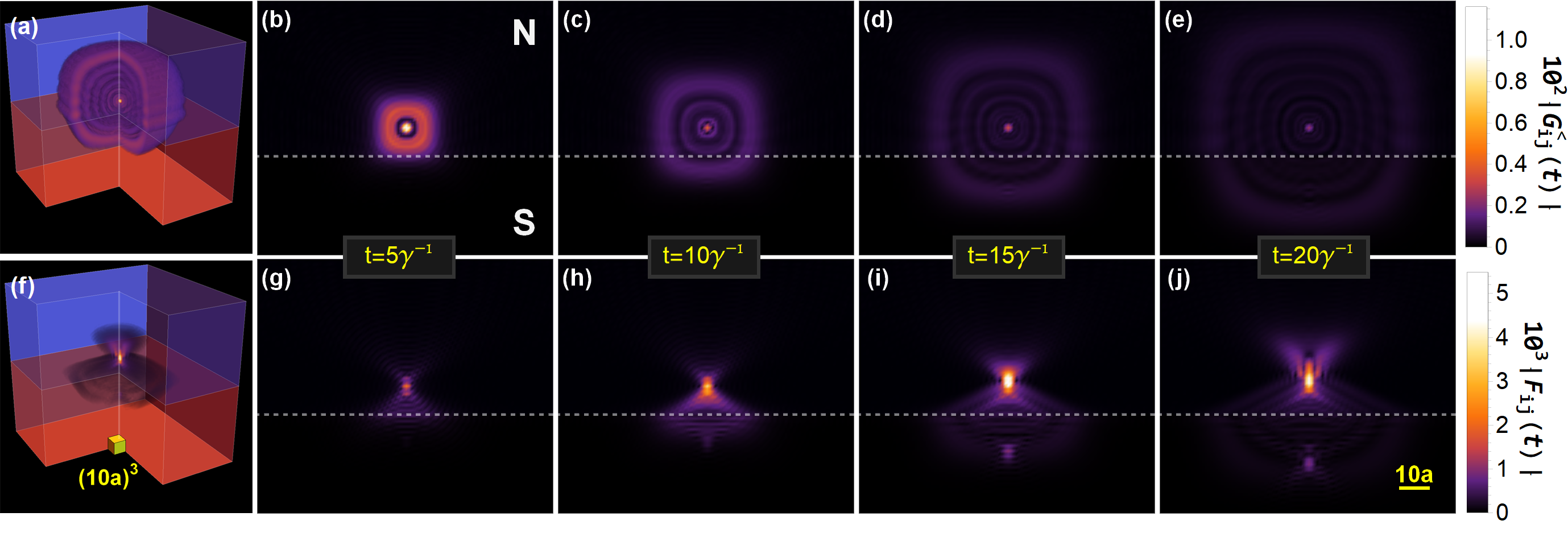}
		\caption{%
		Snapshots in time of the evolution of both the particle-particle $|\mathcal{G}^{<}_{ij}(t)|$ (a-e) and the particle-hole $|\mathcal{F}_{ij}(t)|$ (f-j)  correlation amplitudes for a particle inserted on the normal side near the interface of a 3D NS junction with interface indicated by dashed white line. Spatial density of $|\mathcal{G}^{<}_{ij}(t)|$ (a) and $|\mathcal{F}_{ij}(t)|$ (f) throughout the 3D volume with the nearest quadrant cut-away after an evolution time of $t=20\gamma^{-1}$ and with the length scale shown by the yellow box.
			(b-e) $|\mathcal{G}^{<}_{ij}(t)|$ for a vertical cross-section passing through the start site $j$ for the four different snapshots in time. 
			(g-j) $|\mathcal{F}_{ij}(t)|$ the same cross-section and time steps as in (b-e). 
			The cubic lattice is of dimensions $L^3=(125a)^3$ and extends beyond the regions shown and the spatial scale in indicated by bar in (j). The evolution was computed using the EPOCH method at zero temperature with $M=1000$ moments applied to the Hamiltonian in Eq.~\eqref{Eq:ModelH} at the chemical potential $\mu = 4\gamma$ and pairing potential $\Delta=0.25 \gamma$ inside the superconductor.
		}
		\label{fig:M3D}
	\end{figure*}
		
	We are here interested in the equilibrium time-evolution of a particle (electron) that is initially created at time $t=0$ close to the SN interface on the N side, and then allowed to propagate in time throughout the whole system. The physics of this evolution is captured by the Green's functions $\mathcal{G}^{<}_{ij}(t)$ and $\mathcal{F}_{ij}(t)$, where the former captures the time-evolution of the intact particle, while the latter represents the amplitude associated with a particle at $t=0$ appearing as a hole at a later time, $t$, i.e.~a pair amplitude.
	Because the Hamiltonian in Eq.~\eqref{Eq:ModelH} is isotropic in spin, we can without loss of generality set 
		$\mathcal{G}^{<}_{ij}(t) 
		= 
		i \langle c_{j \sigma}^{\dagger} (0) c_{i \sigma} (t) \rangle	
		$ 
	to be between equal spins and the anomalous correlations are those of a spin-singlet superconducting state between opposite spins 
		$\mathcal{F}_{ij}(t) = i \langle c_{j \downarrow} (0) c_{i \uparrow} (t)  \rangle
		$. 
	We show below that the underlying physics is the same in both 2D and 3D, but because the results are easier to illustrate in 2D we treat this case first.

	In Fig.~\ref{fig:M2D} we show four snapshots in time of the evolution of both $|\mathcal{G}^{<}_{ij}(t)|$ (top row) and $|\mathcal{F}_{ij}(t)|$ (bottom row) for a 2D NS junction of dimensions $L^2=(201a)^2$. The dashed white line marks the interface with the N side on top. Because the superconducting gap also extends into the normal metal through the proximity effect, we can compute the time-evolution using the EPOCH method even at zero temperature to high accuracy, here using a total of $M=1000$ moments.

	As seen in Fig.~\ref{fig:M2D}(a-d), the particle  propagates outwards in approximately circular waves with some anisotropy due to the the anisotropy of the Fermi surface at $\mu = 2 \gamma$. As seen from the last two snapshots in Fig.~\ref{fig:M2D}(c-d) the propagation is largely unimpeded and unaffected even into the superconductor, but on the normal side there are clear interference patterns emanating  parallel to the interface from a partially reflected wave at the SN interface. 
	Studying the anomalous Green's function $|\mathcal{F}_{ij}(t)|$, we see that even at the earliest snap-shot in Fig.~\ref{fig:M2D}(e) there are finite particle-hole correlations on the normal side of the interface, thus appearing even before the main wave of the particle has reached the SN interface. This is a manifestation of the superconducting proximity effect. 
	In the next snap-shot, the particle wave in Fig.~\ref{fig:M2D}(b) 
	has reached the interface where it is transmitted but also Andreev reflected back in the N region as a hole. 
	This shows up in the anomalous part in Fig.~\ref{fig:M2D}(f)
	as growing hole correlations at and beyond the interface.
	As the particle wave further propagates and extends along the SN interface, a part is transmitted as a hole into the S region due to the finite order parameter $\Delta$, resulting in $|\mathcal{G}^{<}_{ij}(t)|$ and $|\mathcal{F}_{ij}(t)|$ showing matching patterns in the S region in Fig.~\ref{fig:M2D}(c,g) and Fig.~\ref{fig:M2D}(d,h). At the SN interface, the intersecting wave crest of $|\mathcal{G}^{<}_{ij}(t)|$ is also Andreev reflected and the correlations propagate back into the N region, producing the triangular hole correlation wave front of $|\mathcal{F}_{ij}(t)|$ in Fig.~\ref{fig:M2D}(h,g) in the N region.

	Because of the $\mathcal{O} (N)$ scaling of the EPOCH method there are no limitations in computing the time-evolved Green's functions also for a 3D bulk system. Thus, similarly to the 2D NS junction, we show in Fig.~\ref{fig:M3D} snapshots of the time-evolution of $|\mathcal{G}^{<}_{ij}(t)|$ and $|\mathcal{F}_{ij}(t)|$ for a 3D NS junction with the system dimensions $L^3=(125a)^3$. The dimension of the underlying BdG Hamiltonian matrix is therefore over $3.9\times 10^6$. Still, using the EPOCH method implemented on a standard laptop we generated Fig.~~\ref{fig:M3D} in a matter of minutes. This clearly illustrates the power and versatility of the EPOCH numerical method.

	Specifically, in Fig.~\ref{fig:M3D}(a,f) we show the spatial extent of $|\mathcal{G}^{<}_{ij}(t)|$ and $|\mathcal{F}_{ij}(t)|$ at $t=20\gamma^{-1}$, respectively throughout the 3D volume surrounding the NS interface resulting from the propagation of a particle inserted on the normal side (blue). To illustrate more details, we also show snapshots at different times of the propagation of $|\mathcal{G}^{<}_{ij}(t)|$ in Fig.~\ref{fig:M3D}(b-e)
	$|\mathcal{F}_{ij}(t)|$ in Fig.~\ref{fig:M3D}(g-j) on a vertical plane intersecting the creation point. 
	From Fig.~\ref{fig:M3D} it is clear that the same physical processes are present in both 2D and 3D. The main difference is only that the amplitudes of $|\mathcal{G}^{<}_{ij}(t)|$ in Fig.~\ref{fig:M3D}(b-e) rapidly diminish on the vertical intersection as the wave's amplitude now spreads out in all three dimensions, consistent with the propagation of spherical waves in 3D. Still, it is possible to discern from Fig.~\ref{fig:M3D}(g-j) that the approximately spherical particle amplitude wave is both Andreev reflected at the interface and hole transformed inside the superconductor,  just as in 2D.

	The results in Figs.~\ref{fig:M2D}-\ref{fig:M3D} illustrate that the EPOCH method readily computes the time-domain Green's functions even for  large 2D and 3D superconducting systems. The method therefore gives direct access to the dynamic equilibrium correlations of any normal or superconducting system, enabling the fully quantum mechanical time-evolution of condensed matter systems to be investigated. Notably, analyzing an NS junction in the time-domain offers a transparency that clearly contributes to an intuitive understanding of the underlying physical processes. Therefore, the EPOCH method provides a new tool for investigating phenomenon occurring in large highly inhomogeneous systems.	
	
\section{Application to linear response}
	\label{sec:LinearResponse}
In this section we highlight yet another potential application of EPOCH, using it in the calculation of correlation functions directly in the time-domain, such as current-current $\langle J_{\mu}(\vec{r},t) J_{\nu}(\vec{r},t) \rangle$ and spin-spin $\langle S_{\mu}(\vec{r},t) S_{\nu}(\vec{r},t) \rangle$  to density-density $\langle n(\vec{r},t) n(\vec{r},t) \rangle$ correlation functions. All these dynamical correlation functions are interesting in their own right and, through the fluctuation-dissipation theorem, they also govern a system's response to external perturbations (not driving the system out of equilibrium), such as magnetic and electric fields \cite{giuliani2005quantum}. For the latter, it is well established that calculations within the Kubo formalism can be carried out using the time-domain equilibrium Green's functions \cite{giuliani2005quantum}, with the susceptibility of an observable $\hat{A} = \sum_{ij} A_{ij} c_{i}^{\dagger} c_{j}$ to a field coupling operator $\hat{B} = \sum_{ij} B_{ij} c_{i}^{\dagger} c_{j}$ being 
			$
				\chi_{AB}(t-t')  = - i \theta(t-t')
						\text{Tr} 
						\left[ 
							A
							\mathcal{G}^{>} (t,t')
							B
							\mathcal{G}^{<} (t',t)
							 -
							A
							\mathcal{G}^{<} (t,t')
							B
							\mathcal{G}^{>} (t',t)
						\right]			
				$.
		Thus the expression for $\chi_{AB}(t-t')$ only requires the Green's functions that are directly computed in EPOCH, given by Eq.~\eqref{Eq:EPPEXP}. As a result, dynamical correlation functions and linear response to external perturbations can be computed efficiently directly in the time-domain using the EPOCH method. Moreover, because the time and temperature dependencies of the response enters only via the transients
		$
				l^{n}_{\beta} ( t ) = (-i)^{n} ( j_{n} ( t)  + i f^{n}_{\beta} ( t ) )
		$, a distinctive advantage of EPOCH is that the response through $\chi_{AB}(t-t')$ can be evaluated at any time or temperature without having to recompute any matrix elements of the quantum propagation. 
		Thus, EPOCH applied to compute the time-domain linear response is more versatile and efficient than previous approaches that rely on a separate finite-difference stepping of the underlying Schr\"odinger equation, which is either only applicable at zero temperature \cite{Iitaka1997} or, if extended to finite temperatures through the Boltzmann-weighted method, requires that the time stepping is redone at each temperature \cite{Iitaka2003}. 
		To conclude, EPOCH is a general method of obtaining the time-domain equilibrium Green's functions in large inhomogeneous systems from which any linear response function can be computed and it automatically gives all temperature and time dependence in a single computation.

In terms of the computational efficiency of EPOCH we note that, with response functions being central to many experiments, they have been a prime target for also other linear scaling methods \cite{Weisse2004,Weise2006}. This is especially the case in the field of quantum transport \cite{Fan2020}, where several new software packages have methods that also allows the conductivity to be computed in large systems \cite{Groth2014, Fan2018, Joao2020, Kloss2020}. Since the conductivity is given by the current-current correlation within the Kubo formula \cite{Greenwood1958, Chester1959, Streda1982, Mayou1988, Mayou1995}, it can also be directly computed using EPOCH.
		In fact, the direct time-domain approach of EPOCH is particularly interesting for quantum transport \cite{Fan2020}, because tracking of wave-packet diffusion in time allows for the identification of ballistic, diffusive, and localized regimes. This explains why the Chebyshev single wave-packet propagation method is still often used in quantum transport works \cite{Roche1997, Roche1999, Ishii2008, Ishii2011, Radchenko2012, Fan2014,PhysRevB.91.165117,Fan2018}, where the conductivity can be computed from the velocity auto-correlation function \cite{Fan2014,Fan2018} or the mean-square displacement \cite{Roche1997,Roche1999,Ishii2008,Ishii2011}. 
		However because single wave-packet propagation does not include the full quantum statistics, unlike EPOCH, any such wave-packet propagation relies on a separate projection on the Fermi energy, as well as normalization by a separately calculated density of states \cite{PhysRevB.91.165117,Fan2020}. With EPOCH we entirely avoid these approximations and separate calculations, because the full quantum statistics is always included at the outset, thus allowing the conductivity tensor to be computed at any temperature and directly in time-domain, from which localization regimes, DC, and AC conductivities can easily be investigated.
		
In relation to the application in quantum transport, we also note that a significant simplification used in many other methods is that the important contributions are usually only from a narrow energy window around the Fermi energy, the so-called Fermi window. This simplification makes it possible to compute the conductivity using only a few low energy states, as e.g.~in the Landauer–Büttiker formalism \cite{Buettiker1985}. Using sparse matrix algorithms, the ($M$) lowest energy states, or generally the scattering states of semi-infinite leads, can then be constructed with a linear scaling ($\mathcal{O}(MN)$) in the system size ($N$) \cite{Santos2019}. One such approach is offered by  the Kwant package \cite{Groth2014}. 
		A further benefit of using wave functions is that external time-dependent perturbations can be incorporated by directly propagating the wave functions using the time-dependent Schr\"odinger equation. Numerically the propagation can be done either with the unitary Crank-Nicolson for stability, or with Runga-Kutta (Linear Multistep Methods) as in the recent TKwant code \cite{Weston2016,Weston2017,Kloss2020}. Thus, for quantum transport problems, an approach based on scattering states is highly efficient, achieving a linear scaling on par with EPOCH. This favorable scaling is however only achieved when considering the narrow low-energy window, and therefore does not extend beyond this ideal transport regime. Since EPOCH always includes all occupied states, EPOCH does not only avoid the cost that comes from working with $M$ wave functions, including computing the necessary overlap integrals and energy integration, but EPOCH can also be directly applied with linear scaling to problems where more than a few occupied states have a finite contribution.

\section{Concluding remarks}
\label{sec:conclusions}
	We develop a computationally efficient method, EPOCH: Equilibrium Propagator by Orthogonal polynomial CHain, to extract the equilibrium thermal time-dependent Green's functions directly in the time-domain that excels for any large inhomogeneous system described by a time-independent Hamiltonian. The method effectively generalizes the widely adopted and successful Chebyshev wave-packet propagation method already widespread in quantum chemistry for single particle pure quantum states to also handle many-body fermionic system in thermal equilibrium. 
	
	The centerpiece of the EPOCH approach is a quantity which we refer to as the equilibrium propagator (EP), whose matrix elements may be used to construct the well-known single particle Green's functions of any fermionic state (such as electron, hole, or Cooper pair) at equilibrium at any fixed temperature. The main advantage of the EPOCH method lies in the efficiency with which this EP may be calculated using an expansion in orthogonal Legendre polynomials. All the time dependence in EPOCH is relegated to the coefficients of these polynomials, the mode transients, which may be written in terms of two parts, a unitary part, simply given by the predefined spherical Bessel functions, and a projective part. The projective transient encodes the quantum statistics by encapsulating all temperature dependence and is readily computed from an efficient recursive relationship that we derive. We arrive at the Green's functions by summing together the mode transients with matrix elements of the polynomial powers of Hamiltonian. These powers are efficiently calculated by single matrix vector multiplications due to the recursive relations of Legendre polynomials. Consequently, the EPOCH method scales linearly in the degrees of freedom of the system, while still both handling time-dependence and equilibrium fermionic statistics. The EPOCH method thus paves the way for investigating quantum phenomenon in large, inhomogeneous, even superconducting, systems, where time-evolution or dynamical aspects are of interest, ranging from quantum transport to odd-frequency superconductivity.
		
	We illustrate the extreme efficiency of the EPOCH method by computing both the normal and anomalous Green's functions for a large 3D superconductor-normal state junction within minutes on a standard laptop. The results show how an electron created on the normal side of the junction propagates towards the interface where it can be either transmitted, normal reflected, or Andreev reflected, and finally builds up a complex pattern of particle-particle (normal) and particle-hole (anomalous or Cooper pair) waves in the junction. We have also recently employed EPOCH to calculate the propagation of even- and odd-frequency pair amplitudes in a disordered normal metal to superconductor junction to uncover unexpected robustness of odd-frequency superconductivity \cite{lothman2020disorder}. 	
	
	In summary, EPOCH is an extremely  computationally effective method to calculate the equilibrium Green's functions directly in the time-domain to capture dynamics in quantum systems, with a wide range of possible applications.
	
\acknowledgements
	We thank  D.~Chakraborty, P.~Dutta, and Y.~Tanaka for helpful discussions and  P.~San-Jos\'{e} for important comments on the manuscript.  We acknowledge financial support from the Swedish Research Council (Vetenskapsr\aa det Grant No.~2018-03488), the Knut and Alice Wallenberg Foundation through the Wallenberg Academy Fellows program, and the European Research Council (ERC) under the European Unions Horizon 2020 research and innovation programme (ERC-2017-StG-757553), and the EU-COST Action CA-16218 Nanocohybri.

\begin{appendix}
\section{Influence of weight functions}\label{app}
		In this appendix we clarify the issue why the weight function $w(x)$ makes the Legendre polynomials a better choice in the EPOCH method than the Chebyshev polynomials.
		Fundamentally, the argument in favor of Legendre polynomials traces back to the fact that a generalized Fourier series of a function $f$ (vector) truncated to the $N$ first terms: $f_N = \sum_{n=0}^{N} \langle e_n, f \rangle e_n$ with respect to an orthogonal basis $\{e_n\}$, is the orthogonal projection onto the subspace spanned by the basis of first basis vectors $\{e_n\}_{n=0}^{N}$. 
		Consequently, the finite series minimizes the residual residue $\left\| f - f_N \right\|$ with respect to the inner product to the function $f$.
		
		In the case of orthogonal polynomials, the inner product is defined via the weight function. 
		Since the weight function of the Legendre polynomials is $w(x)=1$, the residual norm is just the integrated square difference between the approximation and the function $\int_{\mathcal{I}} (f(x) - F_{N}(x) )^2 \dd x $. However, for any other weight function $w(x)$ and polynomial set, the minimized quantity $\int_{\mathcal{I}} w(x) (f(x) - F_{N}(x))^2 \dd x $ necessarily shifts priority from one region to another through $w(x)$, which leads to unequal, and unphysical, treatment of different regions. For example, the weight function of the Chebyshev polynomials is $1/\sqrt{1-x^2} = 1/\sqrt{(1-x)(1+x)}$ and thus has diverging poles at the end of the interval $\mathcal{I}=[-1,1]$. 
		A finite generalized Fourier series in the Chebyshev polynomials therefore 
		sacrifices representational accuracy throughout the interval for a tighter fit around the end points, where the weight function adds most of the cost to any deviation. 
				
		A more quantitative statement of this weight function effect can be made if the generalized Fourier series is combined with the Jackson kernel, introduced in Sec.~\ref{sec:Gibbs}. 
		In the Jackson resummation, the truncated Kernel of Eq.~\eqref{Eq:kernel} is everywhere positive, approximating a Gaussian curve that narrows in on the ideal Dirac delta function, as the number of terms are increased. At the center of the interval $\mathcal{I}=[-1,1]$ i.e.~at $E=0$, which is arguably the most important part of the spectrum of a condensed matter system, 
		the width of this Gaussian kernel is $\pi/N$ using $N$ Chebyshev polynomials \cite{Weise2006}, but only $\pi \sqrt{2/3}/N$ when using $N$ Legendre polynomials. But even more importantly in time-evolution the whole spectrum contributes. Therefore, the main advantage of the  Legendre polynomials is that they avoid placing any arbitrary extra weight to any particular region of the spectrum at the expense of an other region. 
\end{appendix}
	
\bibliographystyle{apsrev4-2}
\bibliography{ODDBIB}

\end{document}